\documentclass[12pt]{article}

\usepackage{amssymb}
\usepackage{citesort}
\usepackage{amsmath}
\usepackage{url}
\usepackage{color}

\newcommand{\Db}{De$\,$Broglie}
\newcommand{\db}{de$\,$Broglie}

\def\la{\langle}
\def\ra{\rangle}
\def\pa{\partial}

\def\vp{\varphi}

\def\be{\begin{equation}}
\def\ee{\end{equation}}
\def\ba{\begin{eqnarray}}
\def\ea{\end{eqnarray}}

\def\bml{\begin{multline}}
\def\eml{\end{multline}}
\def\bal{\begin{align}}
\def\eal{\end{align}}

\begin{document}
\vspace*{1.0cm}
\noindent
{\bf
{\large
\begin{center}
A Dirac sea pilot-wave model for quantum field theory
\end{center}
}
}

\vspace*{.5cm}
\begin{center}
S.\ Colin, W.\ Struyve  \\
Perimeter Institute for Theoretical Physics \\
31 Caroline Street North, Waterloo, Ontario N2L 2Y5, Canada \\
E--mail: scolin@perimeterinstitute.ca, wstruyve@perimeterinstitute.ca
\end{center}

\begin{abstract}
\noindent
We present a pilot-wave model for quantum field theory in which the Dirac sea is taken seriously. The model ascribes particle trajectories to all the fermions, including the fermions filling the Dirac sea. The model is deterministic and applies to the regime in which fermion number is superselected. This work is a further elaboration of work by Colin, in which a Dirac sea pilot-wave model is presented for quantum electrodynamics. We extend his work to non-electromagnetic interactions, we discuss a cut-off regularization of the pilot-wave model and study how it reproduces the standard quantum predictions. The Dirac sea pilot-wave model can be seen as a possible continuum generalization of a lattice model by Bell. It can also be seen as a development and generalization of the ideas by Bohm, Hiley and Kaloyerou, who also suggested the use of the Dirac sea for the development of a pilot-wave model for quantum electrodynamics.
\end{abstract}

\bibliographystyle{unsrt}
\section{Introduction}
In the pilot-wave interpretation of \db\ and Bohm for non-relativistic quantum theory \cite{debroglie28,bohm521,bohm522}, also known as Bohmian mechanics, systems are described by both their wavefunction and by particle positions. The particles move along deterministic trajectories, under the influence of the wavefunction. The incorporation of particles in the description of quantum systems makes it possible to provide a description of reality in which rather ambiguous notions such as `measurement' and `observer' play no fundamental role. 

In this paper we present a way of formulating a pilot-wave theory for quantum fields. In order to put the model into context, we first briefly review the history and status of pilot-wave theories for quantum field theory.

The history of extending non-relativistic pilot-wave theory to quantum fields starts with Bohm's seminal paper \cite{bohm522}. In that paper Bohm gives a pilot-wave interpretation for the quantized electromagnetic field. In this interpretation the additional variables, also called beables, which together with the quantum state make up the description of the quantum field, are not particle positions. Instead the quantum field is also described by an actual field configuration. In a similar way one can construct a pilot-wave interpretation for various other bosonic quantum fields, like for example the scalar field, see \cite{kaloyerou85,kaloyerou94,kaloyerou96,valentini92,valentini96,bohm84,bohm872,bohm93,holland,holland93,lam941,lam942,struyve05,struyve071}. 

In contrast to what might be expected, the introduction of field beables for fermionic quantum fields does not appear to be so straightforward. There is an attempt by Holland \cite{holland,holland881}, who took a field of angular variables as the beable and an attempt by Valentini \cite{valentini92,valentini96}, who took anti-commuting fields as beables. However, in the case of Holland's model some questions need to be addressed before it can be considered a viable model \cite{struyve071}. Valentini's pilot-wave model on the other hand is much more problematic. Although it might seem natural to introduce anti-commuting fields as beables for fermionic quantum fields, it is unclear how this could be done in a way such that the standard quantum predictions are reproduced \cite{struyve05,struyve071}.

Inspired by these difficulties, Struyve and Westman \cite{struyve06} proposed not to introduce beables at all for fermionic quantum fields, only beables for the electromagnetic field. This forms a minimalist pilot-wave interpretation, which is nevertheless capable of reproducing the standard quantum predictions.

A different approach consists of introducing particle positions as beables. This approach has appeared to be more successful for fermionic quantum fields. Bell presented a model for quantum field theory on a lattice \cite{bell872}, where the beables are the fermion numbers at each lattice point. Bell's model differs from the usual pilot-wave program in the fact that it is indeterministic. However, Bell expected that the indeterminism would disappear in the continuum limit. 

Sudbery \cite{sudbery87} was the first to consider a continuum generalization of Bell's model. He considered a non-relativistic particle, without the possibility of particle creation or annihilation, and found that the equation of motion for the particle reduced to the one originally proposed by \db\ and Bohm, which is deterministic. This result was later also obtained by Vink \cite{vink93}. Much later, two different possible continuum generalizations were presented that were capable of dealing with particle creation or annihilation. There is the work by D\"urr {\em et al.}\ \cite{durr02,durr031,durr032,tumulka03,durr04} in which a stochastic generalization is presented and the work by Colin \cite{colin031,colin032,colin033} which deals with a deterministic model.{\footnote{Bell's original model presupposes that space is discrete whereas time is continuous. The possibility of generalizing Bell's model for discrete time has been studied in \cite{barrett05}.}}

The fact that there are two generalizations of Bell's lattice model for the continuum originates in a different reading of Bell's work. D\"urr {\em et al.}\ take the fermion number to be the number of particles plus the number of anti-particles. Their model therefore ascribes trajectories to both particles and anti-particles. The trajectories are deterministic until there is particle creation or annihilation.

Colin has a different reading of fermion number. In standard quantum field theory the `fermion number' is often taken to mean the number of particles minus the number of anti-particles. We call this number $F$. Instead of using the particle--anti-particle picture one can also use the Dirac sea picture, in which an anti-particle is a hole in the sea of negative energy particles. In the Dirac sea picture of quantum field theory, which is generally considered to be equivalent to the particle--anti-particle picture, there is a natural notion of the fermion number, namely the number of positive plus negative energy particles. We call this number $F_d$. This number just equals $F$ plus the number of particles in the Dirac sea, the latter being an infinite constant. We keep calling both $F$ and $F_d$ the fermion number. Colin now takes the fermion number to mean $F_d$.

So, in the work by Colin, which deals with quantum electrodynamics, the picture of a Dirac sea is taken seriously. Position beables are introduced for the particles in the Dirac sea, i.e.\ the negative energy particles, and for the positive energy particles. It is the fact that the fermion number $F_d$, as the number of positive plus negative energy particles, is always positive, together with the fact that it can be considered superselected in quantum electrodynamics, that allows the construction of a deterministic model in a similar fashion as in non-relativistic pilot-wave theory. Fermion number superselection means that the quantum state cannot evolve into a superposition of states with different fermion number. In particular superselection of fermion number implies conservation of fermion number, i.e.\ that the fermion number operator commutes with the Hamiltonian. (Note that fermion number conservation does not imply fermion number superselection. Although conservation of fermion number implies that the fermion number of a quantum state will never change during Schr\"odinger evolution, it could still change during collapse, in which case the fermion number would not be superselected.)  

The model that Colin found as a continuum version of Bell's model for lattice quantum field theory is a model that was suggested earlier, though not developed in detail, by Bohm, Hiley and Kaloyerou, see \cite{bohm872} and \cite[p.\ 276]{bohm93}. In this paper, we further elaborate on this model. We study to what extent the model can be generalized as to account for other types of interactions, such as the weak and strong interaction. In addition, we study how the pilot-wave model can be regularized, i.e.\ how it can be made mathematically well defined. The regularization is required at two levels, namely at the level of the Schr\"odinger equation which describes the evolution of the quantum state, and at the level of the equations describing the beables. Instead of introducing a lattice regularization, like Bell, we introduce a cut-off regularization. This allows us to keep continuous space and to preserve the determinism of the pilot-wave model. We also study how the pilot-wave model reproduces the predictions of standard quantum field theory.

As mentioned before, a key ingredient in the construction of the Dirac sea pilot-wave model of Colin and Bohm {\em et al.}\ is that the fermion number can be considered superselected in the case of electromagnetic interactions. In particular, the fermion number is conserved. To date, no experiment has shown a violation of fermion number conservation, not even if non-electromagnetic interactions are involved \cite{yao06}. Moreover, all Feynman diagrammatic processes in the Standard Model preserve the fermion number. A familiar example is the weak process $d \to u + e + {\bar \nu}_e$, which takes place in $\beta$-decay. Nevertheless the Standard Model does predict a violation of fermion number conservation. This predicted violation of the fermion number is a non-perturbative effect, i.e.\ it does not appear in perturbative treatments, such as Feynman diagrammatic ones. The violation is a prediction of the electroweak theory \cite{thooft76} and only appears for sufficiently high energies. According to one recent estimate \cite{ringwald03}, a violation is expected to be observed no sooner than in the Very Large Hadron Collider (the hypothetical future hadron collider with performance significantly beyond the Large Hadron Collider). 

Our generalization of the Dirac sea pilot-wave model of Colin and Bohm {\em et al.}\ will hold in the regime where the fermion number is conserved. Although the pilot-wave model can therefore not reproduce all the predictions of the Standard Model, it can at least reproduce the predictions of the Standard Model in some low energy regime. For example the pilot-wave model can perfectly describe the regime in which the electro-weak symmetry is broken by spontaneous symmetry breaking. In this paper, we will therefore restrict our attention to this energy regime. The model of D\"urr {\em et al.}\ does not rely on the conservation of any quantum number and therefore seems capable of describing fermion number violating interactions. 

Next to fermion number superselection, also the introduction of a Dirac sea is a key ingredient in the construction of the pilot-wave model. In a filled Dirac sea all negative energy states, corresponding to quarks and leptons with all possible quantum numbers, are occupied. Superselection and positivity of the fermion number $F_d$ then imply that we can introduce position beables in a similar way as in non-relativistic quantum mechanics. The position beables are further not distinguished by any label corresponding to properties such as mass, spin, charge or flavour. This is similar to Bell's model for lattice quantum field theory, since the beables in this model are the numbers of fermions at each lattice point.

With the introduction of position beables also for the negative energy fermions, the Dirac sea pilot-wave model yields an ontology which is not immediately related to our image of a classical world. Furthermore, in standard quantum field theory measurements are not expressed in terms of fermion number densities which include the negative energy fermion contributions. As a result, it is not so straightforward to show that the pilot-wave model reproduces the predictions of standard quantum field theory. Instead of providing conclusive evidence, we only give arguments that support our expectation that the model reproduces the standard quantum predictions.

We also want to mention that the pilot-wave model is not Lorentz covariant at the fundamental level. However, insofar as the model reproduces the predictions of standard quantum field theory, Lorentz covariance is regained at the statistical level. There exist some attempts to construct a Lorentz covariant pilot-wave model, see \cite{tumulka06} for a good review. On the other hand, see also \cite{valentini97} for an argument that the fundamental symmetry of pilot-wave theories is Aristotelian rather than Lorentzian.

The outline of the paper is as follows. First, in Section \ref{general framework} we describe in general the technique by which the pilot-wave model is constructed. Then in Section \ref{the model}, the actual model is presented in the context of the effective quantum field theory which arises in the Standard Model after spontaneous symmetry breaking. In Section \ref{ultravioletcutoff}, we discuss a cut-off regularization and indicate how the model reproduces the standard quantum field theoretical predictions. Finally, in Section \ref{violationfermionnumber} we add a note on the predicted violation of fermion number in the standard model.

\section{General framework}\label{general framework}
Before we pass to the general framework for constructing the pilot-wave model, we first recall the pilot-wave interpretation for non-relativistic quantum theory that was originally proposed by \db\ and Bohm \cite{debroglie28,bohm521,bohm522}. In this interpretation, a quantum system is described by its wavefunction $\psi({\bf x}_1,\dots,{\bf x}_n,t)$ in configuration space ${\mathbb R}^{3n}$ for all times $t$, which satisfies the non-relativistic Schr\"odinger equation
\begin{equation}
i \hbar \frac{\partial \psi({\bf x}_1, \dots,{\bf x}_n,t)}{\pa t} =  \left( -\sum^n_{k=1} \frac{\hbar^2 \nabla^2_k}{2m_k} + V({\bf x}_1, \dots,{\bf x}_n) \right)  \psi({\bf x}_1, \dots,{\bf x}_n,t)\,,
\label{1}
\end{equation}
and by $n$ particles, which move in physical space ${\mathbb R}^3$, and for which the possible trajectories ${\bf x}_k(t)$ are solutions to the guidance equations
\begin{equation}
\frac{{d} {\bf x}_k}{dt} = {\bf{v}}^{\psi}_k = \frac{ {\boldsymbol{\nabla}}_k S }{m_k} \,, 
\label{2}
\end{equation}
where $\psi=|\psi|\exp{iS/\hbar}$. 

How the pilot-wave theory reproduces the standard predictions of quantum theory has been discussed before in many places, see e.g.\ \cite{bohm522}. A key ingredient is that the positions of the particles are distributed according to $|\psi|^2$ over ensembles where all the systems have wavefunction $\psi$. The special distribution $|\psi|^2$ is called the equilibrium distribution \cite{valentini911,durr92}. The particular form of the guidance equations guarantees that if the distribution of particle positions equals $|\psi({\bf x}_1, \dots,{\bf x}_n,t_0)|^2$ at a certain time $t_0$, then it will equal $|\psi({\bf x}_1, \dots,{\bf x}_n,t)|^2$ at times $t$. This means that the distribution keeps its functional form of $\psi$. This property is also called equivariance \cite{durr92}. Possible justifications of quantum equilibrium have been discussed in \cite{valentini911,valentini92,durr92,bohm93}.

There is a simple way to derive the pilot-wave interpretation. Our Dirac sea pilot-wave model for quantum field theory will be derived along similar lines. This derivation goes as follows. In standard non-relativistic quantum theory $|\psi({\bf x}_1, \dots,{\bf x}_n,t)|^2$ is interpreted as the probability density to find the particles in the respective volumes $d^3x_i$ around the positions ${\bf x}_i$, in a measurement on an ensemble of systems all described by the wavefunction $\psi$. The conservation of probability yields the continuity equation 
\begin{equation}
\frac{\partial |\psi|^2}{\partial t} + \sum^n_{k=1} {\boldsymbol{\nabla}}_k \cdot \left( {\bf v}^{\psi}_k |\psi|^2 \right) =0\,,
\label{3}
\end{equation}
which can be derived from the Schr\"odinger equation (\ref{1}). If one insists that particles have positions at all times, i.e.\ also if no measurement is performed, then the analogy with fluid dynamics suggests the equation of motion (\ref{2}) for the particles. In this way we arrive at the pilot-wave interpretation. 

Note that originally, neither \db\ nor Bohm followed the above derivation to arrive at the pilot-wave theory \cite{bacciagaluppi06}. \Db\ postulated a precursor of the guidance equation before the wave equation was known, so that the guidance equation was not obtained from considering the wave equation first. In his Ph.D.\ thesis in 1924 \db\ namely postulated that the momenta of particles should be proportional to the gradient of a phase, where the phase corresponds to an unspecified periodic phenomenon. Only later, in 1926,  Schr\"odinger provided a wave equation for this phase. On the other hand, when Bohm rediscovered the pilot-wave theory in 1951, he was in the first place motivated by the resemblance to classical Hamilton-Jacobi theory, which arises when the Schr\"odinger equation is rewritten in terms of the polar decomposition of the wavefunction.

Nevertheless, this way of deriving the pilot-wave theory provides a scheme which can be used (and is often used) to construct pilot-wave models for other quantum theories. Let us sketch this scheme. We restrict ourselves to the case in which particle positions are introduced as beables. The scheme is immediately amenable for the introduction of other type of beables, like fields, strings, etc.

Suppose we have a Schr\"odinger equation 
\begin{equation}
i\hbar\frac{ d  \left| \psi (t)\right\rangle }{dt} = {\widehat H} \left| \psi(t) \right\rangle
\label{4}
\end{equation}
with Hamiltonian ${\widehat H}$ and a positive operator valued measure (POVM) ${\widehat P}(d^3 x_1 \dots d^3 x_n)$ for position. This means that $\left\langle \psi(t) \right|  {\widehat P}(d^3 x_1 \dots d^3 x_n) \left| \psi(t) \right\rangle$ gives the probability to find $n$ particles in the respective volumes $d^3x_i$ around the positions ${\bf x}_i$ in a measurement at time $t$.{\footnote{For example, in the case of non-relativistic quantum theory we have a position PVM, i.e.\ a projection valued measure, which is a special case of a POVM. The position PVM is given by 
\begin{equation}
{\widehat P}(d^3 x_1 \dots d^3 x_n) = \left|{\bf x}_1, \dots,{\bf x}_n \right\rangle \left\langle {\bf x}_1, \dots,{\bf x}_n \right| d^3 x_1 \dots d^3 x_n\,,
\label{5.1}
\end{equation}
where the states $\left|{\bf x}_1, \dots,{\bf x}_n \right\rangle$ are the standard position eigenstates.}}  This probability determines a probability density $\rho^{\psi}({\bf x}_1, \dots,{\bf x}_n,t)$ by the relation
\begin{equation}
\rho^{\psi}({\bf x}_1, \dots,{\bf x}_n,t) d^3 x_1 \dots d^3 x_n = \left\langle \psi(t) \right|  {\widehat P}(d^3 x_1 \dots d^3x_n) \left| \psi(t) \right\rangle\,.
\label{5}
\end{equation}
The Schr\"odinger equation implies the following time evolution for the probability density: 
\begin{equation}
\frac{\partial \rho^{\psi}({\bf x}_1, \dots,{\bf x}_n,t)}{\partial t} d^3 x_1 \dots d^3 x_n +  \left\langle \psi(t) \right| i\hbar\left[  {\widehat P}(d^3 x_1 \dots d^3x_n) , {\widehat H}  \right]\left| \psi(t) \right\rangle=0\,.
\label{6}
\end{equation}
If we can write the second term in the form 
\begin{equation}
\left\langle \psi(t) \right| i\hbar\left[  {\widehat P}(d^3 x_1 \dots d^3x_n) , {\widehat H}  \right]\left| \psi(t) \right\rangle =  \sum^n_{k=1} {\boldsymbol{\nabla}}_k  \cdot \left( {\bf v}^{\psi}_k \rho^{\psi}({\bf x}_1, \dots,{\bf x}_n,t) \right) d^3 x_1 \dots d^3x_n\,,
\label{7}
\end{equation}
where the ${\bf v}^{\psi}_k$ are some vector fields, then we can construct a pilot-wave model with particle positions as beables, by postulating the guidance equations
\begin{equation}
\frac{{d} {\bf x}_k}{dt} = {\bf{v}}^{\psi}_k\,.
\label{8}
\end{equation}
The equilibrium density for the particle positions is given by $\rho^{\psi}$.

Note that we can not always write the second term in (\ref{6}) in the form (\ref{7}), for example when the number of particles is not conserved. In such cases one still has the pilot-wave-type models of D\"urr {\em et al.}\ which include stochastic jumps \cite{durr02,durr031,durr032,tumulka03,durr04}.

\section{The pilot-wave model}\label{the model} 
In this section, we present our pilot-wave model for quantum field theory. We start with the Hamiltonian of the Standard Model after spontaneous symmetry breaking, so that the weak interaction bosons and the fermions have acquired their mass. This means that we do not have to worry about the predicted violation of fermion number conservation, which only appears for higher energies. We further discuss the predicted violation of fermion number in Section \ref{violationfermionnumber}. 

At this stage, we do not introduce a regularization of the theory. We postpone the introduction of a regularization to Section \ref{ultravioletcutoff}.

In this and the following sections we use units in which $\hbar=c=1$.

\subsection{The Hamiltonian}
The Hamiltonian is of the form
\begin{equation}
\widehat{H}=\widehat{H}^B + \widehat{H}^F_0+\widehat{H}_I\,.
\label{9}
\end{equation}
The part $\widehat{H}^B$ is the part of the Hamiltonian that contains only bosonic fields, i.e.\ the interaction fields and the Higgs field. It contains the free Hamiltonian of the bosonic fields together with all the coupling terms between the various bosonic fields. The part $\widehat{H}^F_0$ is the free Hamiltonian for the fermionic fields, i.e.\ the quark and lepton fields. $\widehat{H}_I$ contains the interaction terms between the fermionic and bosonic fields.

The Hamiltonian depends on the Dirac field operators $\widehat{\psi}_\lambda({\bf x})$ of the quarks and leptons. The label $\lambda$ comprises quantum numbers like lepton number, quark number, quark colour, etc.{\footnote{Note that we take the neutrinos to be Dirac particles. The possibility still exists that neutrinos are Majorana particles. In that case the neutrino would be its own anti-particle, which would lead to a violation of fermion number conservation.}} The Dirac fields $\widehat{\psi}$ and their conjugates $\widehat{\psi}^\dagger$ satisfy the equal-time anti-commutation relations 
\begin{equation}
\{\widehat{\psi}_{\lambda,a}({\bf x}),\widehat{\psi}^\dagger_{\lambda',a'}({\bf x}')\}=\delta_{\lambda{\lambda'}}\delta_{aa'}\delta({\bf x}-{\bf x}')\,.
\label{9.1}
\end{equation}
The other fundamental anti-commutation relations are zero. The labels $a$ and $a'$ are Dirac spinor indices. We give a representation for these operators in the following subsection. 

For the bosonic operators we do not introduce an explicit quantization scheme. We leave the corresponding bosonic Hilbert space and its orthonormal basis $\left\{ | \xi \ra \right\}$ unspecified. We expect that the subtleties in quantizing the bosonic fields do not affect the key properties of our pilot-wave model, at least not in the presently considered energy regime.

The free Hamiltonian for the fermionic fields $\widehat{H}^F_0$ and the interaction Hamiltonian $\widehat{H}_I$ are respectively given by
\begin{eqnarray}
\widehat{H}^F_0 &=& \sum_{\lambda} \int d^3x \widehat{\psi}^{\dagger}_\lambda({\bf x})[ -i {\boldsymbol \alpha}\cdot{\boldsymbol \nabla} + \beta m_\lambda] \widehat{\psi}_\lambda({\bf x})\,, \\
\widehat{H}_I &=& \sum_{\lambda,\lambda',a,a'} \int d^3x \widehat{\psi}^{\dagger}_{\lambda,a} ({\bf x})  {\widehat h}_{\lambda,a,\lambda',a'}({\bf x}) \widehat{\psi}_{\lambda',a'}({\bf x})\,,
\label{10}
\end{eqnarray}
where $m_\lambda$ is the mass of the particle with quantum numbers $\lambda$. The kernel ${\widehat h}_{\lambda,a,\lambda',a'}({\bf x})$ carries a hat since it involves the bosonic field operators. It contains no fermionic field operators. The form of the interaction Hamiltonian is typical for an interaction introduced by minimally coupling the free fermionic Hamiltonian $\widehat{H}^F_0$ with gauge fields. Also the Yukawa interaction which couples the fermionic fields to the Higgs boson is of this form.

Instead of writing out the bosonic Hamiltonian $\widehat{H}^B$ or the interaction Hamiltonian $\widehat{H}_I$ explicitly, let us just present parts of the latter here. For the electromagnetic interaction Hamiltonian, for example, the kernel is given by
\begin{equation}
{\widehat h}_{\lambda,a,\lambda',a'}({\bf x}) = q_\lambda \delta_{\lambda \lambda'} \left( \gamma^0 \gamma^\mu \right)_{aa'}{\widehat A}_\mu({\bf x})\,,
\label{10.1}
\end{equation}
where $q_\lambda$ is the charge of the particle with quantum numbers $\lambda$ and ${\widehat A}_\mu$ the vector potential. Another example is the interaction Hamiltonian which represents the coupling of the leptons to the weak interaction vector bosons $\widehat{W}^{\pm}_\mu$. In order to have a simple presentation of this Hamiltonian let $\lambda$ stand for just the lepton flavour. We write $\lambda=(l,s)$, where $l$ represents the lepton family, i.e.\ electron, muon or tau family, and where $s=1,2$ indicates the electrically charged or neutral lepton (i.e.\ the neutrino) in the family. With this notation the kernel reads
\begin{equation}
\widehat{h}_{\lambda,a,\lambda',a'}({\bf x})=\frac{g}{2\sqrt{2}}\delta_{ll'}\left(\gamma^0\gamma^\mu(1-\gamma^5)\right)_{a a'}\left(\delta_{s2}\delta_{s'1}\widehat{W}^-_\mu({\bf x})+\delta_{s1}\delta_{s'2} \widehat{W}^+_\mu({\bf x})\right)\,,
\label{10.2}
\end{equation}
where $g$ is the coupling constant for the group $SU(2)$ in the electroweak theory.

\subsection{The Dirac sea picture}
The standard Fourier expansion of the Dirac field operator is given by
\begin{equation}
\widehat{\psi}_\lambda({\bf x})=\sum_{s}\int \frac{d^3 p}{\sqrt{(2\pi)^3}} \sqrt{\frac{m_\lambda}{E_\lambda({\bf p})}} \left[\widehat{c}_{\lambda,s}({\bf p})u_{\lambda,s}({\bf p})e^{i{\bf p}\cdot{\bf x}} + \widehat{d}^\dagger_{\lambda,s}({\bf p})v_{\lambda,s}({\bf p})e^{-i{\bf p}\cdot{\bf x}} \right]\,,
\label{11}
\end{equation} 
where $E_\lambda({\bf p})=\sqrt{ {\bf p}\cdot{\bf p} +m^2_\lambda}$. The operators $\widehat{c}^\dagger_{\lambda,s}$ and $\widehat{c}_{\lambda,s}$ respectively create and annihilate particles with quantum numbers $\lambda$ and helicity $s$. The operators $\widehat{d}^\dagger_{\lambda,s}$ and $\widehat{d}_{\lambda,s}$ respectively create and annihilate the corresponding anti-particles. They satisfy the relations 
\begin{equation}
\{\widehat{c}_{\lambda,s}({\bf p}),\widehat{c}^\dagger_{\lambda',s'}({\bf p}')\}=\delta_{\lambda\lambda'}\delta_{ss'}\delta({\bf p}-{\bf p}')\,, \qquad \{\widehat{d}_{\lambda,s}({\bf p}),\widehat{d}^\dagger_{\lambda',s'}({\bf p}')\}=\delta_{\lambda\lambda'}\delta_{ss'}\delta({\bf p}-{\bf p}')\,.
\label{11.1}
\end{equation}
The remaining fundamental anti-commutation relations are zero.

Instead of expressing the Dirac field operators in terms of creation and annihilation operators for particles and anti-particles, which have positive energy, we can express them in terms of creation and annihilation operators for particles of positive and negative energy 
\begin{equation}
{\widehat \psi}_\lambda({\bf x})=\sum_{s}\int \frac{d^3 p}{\sqrt{(2\pi)^3}} \sqrt{\frac{m_\lambda}{E_\lambda({\bf p})}} \left[\widehat{c}_{\lambda,s}({\bf p})u_{\lambda,s}({\bf p})e^{i{\bf p}\cdot{\bf x}} + \widehat{\zeta}_{\lambda,s}({\bf p})v_{\lambda,s}(-{\bf p})e^{i{\bf p}\cdot{\bf x}} \right]\,.
\label{12}
\end{equation} 
The operators $\widehat{\zeta}^\dagger_{\lambda,s}$ and $\widehat{\zeta}_{\lambda,s}$ respectively create and annihilate particles with negative energy. They are connected to the anti-particle operators by the relations $\widehat{\zeta}_{\lambda,s}({\bf p})=\widehat{d}^\dagger_{\lambda,s}(-{\bf p})$.

In standard quantum field theory the state containing no particles or anti-particles is noted as $|0\ra$. In the absence of any interactions this state would be the vacuum. In the Dirac sea picture this state is obtained from the Dirac sea vacuum $|0_D\ra$, which is the state that contains no particles of positive or negative energy, by applying all the creation operators of negative energy particles, i.e.
\begin{equation}
|0\ra = \prod_{\lambda,s,{\bf p}} {\widehat \zeta}^\dagger_{\lambda,s}({\bf p}) |0_D\ra \,.
\label{12.1}
\end{equation} 
In this picture, an anti-particle corresponds to a hole in the Dirac sea.

Historically, the Dirac sea picture was convenient to interpret the Dirac equation as a many particle equation that was able to explain the appearance of pair creation in quantum electrodynamics. For non-electromagnetic interaction the Dirac sea picture was abandoned in favour of standard quantum field theory \cite{sakurai67}. In standard quantum field theory particles and anti-particles can be freely created, without the need to refer to the Dirac sea picture. However, in order to construct our pilot-wave model it will appear advantageous to reintroduce the Dirac sea picture. It will allow us to formulate a pilot-wave theory in which particles travel along deterministic trajectories.

\subsection{The fermion number operator and position PVM}\label{fn}
The fermion number operator is given by 
\begin{equation}
\widehat{F}_d=\int d^3x\widehat{F}_d({\bf x})\,,
\label{12.2}
\end{equation}
with 
\begin{equation}
\widehat{F}_d({\bf x})=\sum_{\lambda,a}\widehat{\psi}^\dagger_{\lambda,a}({\bf x})\widehat{\psi}_{\lambda,a}({\bf x})
\label{13}
\end{equation}
the fermion number density operator. The fact that fermion number indeed corresponds to the total number of positive and negative energy particles, or equivalently the total number of particles minus the number of anti-particles plus an infinite constant, can be seen from the expression
\begin{eqnarray}
\widehat{F}_d &=& \int d^3x \widehat{F}_d({\bf x}) \nonumber \\
&=& \sum_{\lambda,s} \int d^3p \left( \widehat{c}^\dagger_{\lambda,s}({\bf p}) \widehat{c}_{\lambda,s}({\bf p}) + \widehat{\zeta}^\dagger_{\lambda,s} ({\bf p})\widehat{\zeta}_{\lambda,s} ({\bf p}) \right) \nonumber \\
&=& \sum_{\lambda,s} \int d^3p \left( \widehat{c}^\dagger_{\lambda,s}({\bf p}) \widehat{c}_{\lambda,s}({\bf p}) +\widehat{d}_{\lambda,s}({\bf p})\widehat{d}^\dagger_{\lambda,s}({\bf p})  \right)\nonumber \\
&=& \sum_{\lambda,s} \int d^3p \left( \widehat{c}^\dagger_{\lambda,s}({\bf p}) \widehat{c}_{\lambda,s}({\bf p}) - \widehat{d}^\dagger_{\lambda,s}({\bf p})\widehat{d}_{\lambda,s}({\bf p})  + \delta({\bf 0})\right)\,.
\label{14}
\end{eqnarray}
The state with zero fermion number is the state $|0_D\rangle$.

Let us now introduce the states
\begin{equation}
\left|{\bf x}_1,\lambda_1,a_1;\dots; {\bf x}_n,\lambda_n,a_n \right\ra = \frac{1}{{\sqrt{n!}}} {\widehat \psi}^{\dagger}_{\lambda_1,a_1}({\bf x}_1) \dots  {\widehat \psi}^{\dagger}_{\lambda_n,a_n}({\bf x}_n)|0_D\rangle\,.
\label{14.2}
\end{equation}
These states are normalized as
\begin{multline}
\la {\bf x}_1,\lambda_1,a_1;\dots; {\bf x}_n,\lambda_n,a_n \left|{\bf x}'_1,\lambda'_1,a'_1;\dots; {\bf x}'_n,\lambda'_n,a'_n \right\ra   \\
 =  \frac{1}{n!}  \sum_\sigma (-1)^\sigma \delta({\bf x}_{\sigma(1)} -  {\bf x}'_1)\delta_{\lambda_{\sigma(1)}\lambda'_1}   \delta_{a_{\sigma(1)}a'_1}  \dots  \delta({\bf x}_{\sigma(n)} - {\bf x}'_n)\delta_{\lambda_{\sigma(n)}\lambda'_n}   \delta_{a_{\sigma(n)}a'_n}\,,   
\label{15}
\end{multline}
where the sum runs over the permutations of the indices $1,\dots,n$ and where $(-1)^\sigma$ is the sign of the permutation. Making use of the commutation relations
\begin{equation}
\left[\widehat{F}_d({\bf y}), \widehat{\psi}^\dagger_{\lambda,a}({\bf x}) \right] =  \widehat{\psi}^\dagger_{\lambda,a}({\bf y}) \delta({\bf x}-{\bf y}) \,,
\label{15.1}
\end{equation}
one can easily show that the states $\left|{\bf x}_1,\lambda_1,a_1;\dots; {\bf x}_n,\lambda_n,a_n \right\ra$ are eigenstates of the fermion number density operator
\begin{equation}
\widehat{F}_d({\bf x}) \left|{\bf x}_1,\lambda_1,a_1;\dots; {\bf x}_n,\lambda_n,a_n \right\ra = \sum^n_{i=1} \delta({\bf x} - {\bf x}_i) \left|{\bf x}_1,\lambda_1,a_1;\dots; {\bf x}_n,\lambda_n,a_n \right\ra\,.
\label{16}
\end{equation}
The corresponding eigenvalues are degenerate since they do not depend on the $\lambda_i$ or $a_i$. 

It is clear that the states $\left|{\bf x}_1,\lambda_1,a_1;\dots; {\bf x}_n,\lambda_n,a_n \right\ra$ represent, for all values of $\lambda_i$ and $a_i$, states for which there is a fermion at the positions ${\bf x}_1$ to ${\bf x}_n$. The total fermion number is equal to $n$. Therefore the PVM
\begin{multline}
{\widehat P}(d^3x_1 \dots d^3x_n)= \sum_{\lambda_1,a_1,\dots,\lambda_n,a_n,\xi}  \left|{\bf x}_1,\lambda_1,a_1;\dots; {\bf x}_n,\lambda_n,a_n  \right\ra \otimes \left| \xi \right\ra \\
\times \left\la \xi \right| \otimes  \left\la{\bf x}_1,\lambda_1,a_1;\dots; {\bf x}_n,\lambda_n,a_n \right| d^3x_1 \dots d^3x_n 
\label{17}
\end{multline}
can be interpreted as a position PVM (remember that the states $| \xi \ra$ form an orthonormal basis of the bosonic Hilbert space). The fact that the measure ${\widehat P}(d^3x_1 \dots d^3x_n)$ is projection operator valued is straightforwardly checked by using the normalization ({\ref{15}}) of the states $\left|{\bf x}_1,\lambda_1,a_1;\dots; {\bf x}_n,\lambda_n,a_n \right\ra$. 

The position PVM ${\widehat P}(d^3x_1 \dots d^3x_n)$ also corresponds to the conventional position operator in the Dirac theory when the Dirac sea picture is adopted, see e.g.\ \cite{barut68}. It is for example for this position operator that Schr\"odinger derived his famous {\em Zitterbewegung} \cite{schrodinger30}. Note that the position eigenstates $\left|{\bf x}_1,\lambda_1,a_1;\dots; {\bf x}_n,\lambda_n,a_n \right\ra$ are superpositions of positive and negative energy states. Therefore these states are sometimes called `abstract' position eigenstates, to contrast them with states which are built out only of positive energy states, like for example the eigenstates of the Newton-Wigner position operator \cite{barut68}.

Now we would like to regard the states $\left|{\bf x}_1,\lambda_1,a_1;\dots; {\bf x}_n,\lambda_n,a_n \right\ra$ with $n\in {\mathbb{N}}$ as a basis for our Hilbert space. However, the Hilbert space spanned by these states would not contain states of physical interest. Physically relevant states, which are states which contain a finite number of positive energy particles and a finite number of holes in the negative energy sea, namely have an infinite fermion number. However, dealing with an infinite number of particles would lead to a number of complications in maintaining mathematical rigour. First of all we would have the field theoretical divergences. Secondly, since there does not exist a generalization of the Lebesgue measure to an infinite dimensional space, one cannot assume the naive infinite particle number limit in expressions like those for the position PVM \cite{tumulka06}. We will come back to these issues in Section \ref{needregularization}. For the moment, we keep working with states with finite fermion number. The hope is that some key properties of the pilot-wave model survive when passing to some regularized theory. In Section \ref{ultravioletcutoff}, we discuss in detail such a regularized theory, which is obtained by introducing an ultra-violet momentum cut-off and by assuming a finite space.

\subsection{Fermion number superselection}
Since the fermion number operator $\widehat{F}_d$ commutes with the Hamiltonian $\widehat{H}$, the fermion number is conserved. We further assume that fermion number is superselected. As explained in the introduction this assumption is empirically adequate, although that may change in the future if the predictions of Standard Model turn out to be correct. The fermion number superselection implies that we can restrict the fermionic Hilbert space, which is spanned by the states $\left|{\bf x}_1,\lambda_1,a_1;\dots; {\bf x}_n,\lambda_n,a_n \right\ra$ with $n\in {\mathbb{N}}$, to a subspace spanned by states $\left|{\bf x}_1,\lambda_1,a_1;\dots; {\bf x}_n,\lambda_n,a_n \right\ra$, where the fermion number $n$ is fixed.

In the Hilbert space with fixed fermion number $n$, we can expand a quantum state $\left| \psi (t)\right\rangle$ as
\begin{multline}
\left| \psi (t)\right\rangle =\sum_{\lambda_1,a_1,\dots,\lambda_n,a_n,\xi}  \int d^3x_1 \dots d^3x_n \\
\psi_{\lambda_1,a_1,\dots,\lambda_n,a_n,\xi}({\bf x}_1,\dots,{\bf x}_n,t) \left|{\bf x}_1,\lambda_1,a_1;\dots; {\bf x}_n,\lambda_n,a_n \right\ra \otimes \left| \xi \right\ra\,, 
\label{18}
\end{multline}
where the expansion coefficients $\psi_{\lambda_1,a_1,\dots,\lambda_n,a_n,\xi}({\bf x}_1,\dots,{\bf x}_n,t)$, which are given by
\begin{equation}
\psi_{\lambda_1,a_1,\dots,\lambda_n,a_n,\xi}({\bf x}_1,\dots,{\bf x}_n,t) =\left(  \left\langle {\bf x}_1,\lambda_1,a_1;\dots; {\bf x}_n,\lambda_n,a_n \right| \otimes \left\langle \xi \right| \right) \left| \psi (t)\right\rangle \,,
\label{18.01}
\end{equation}
are anti-symmetric under permutations of the indices $1,\dots,n$. In terms of these expansion coefficients, the position density that corresponds to the position PVM ({\ref{17}}) is given by, cf.\ equation ({\ref{5}}),
\begin{equation}
\rho^{\psi}({\bf x}_1, \dots,{\bf x}_n,t) = \sum_{\lambda_1,a_1,\dots,\lambda_n,a_n,\xi} \left| \psi_{\lambda_1,a_1,\dots,\lambda_n,a_n,\xi}({\bf x}_1,\dots,{\bf x}_n,t) \right|^2\,.
\label{18.1}
\end{equation}

Since we have a fixed fermion number, the fermion number density operator can be written as 
\begin{equation}
\widehat{F}_d({\bf x}) = \int \sum^n_{i=1} \delta({\bf x} - {\bf x}_i) {\widehat P}(d^3x_1 \dots d^3x_n)\,.
\label{18.2}
\end{equation}
For later purposes, we also introduce the operator
\begin{equation}
\widehat{F}_d(B) = \int_B d^3x \widehat{F}_d({\bf x}) \,,
\label{19}
\end{equation}
which is the operator corresponding to the number of fermions in the region $B$. The associated expectation value $\left\la \psi(t) \right| \widehat{F}_d(B) \left| \psi(t) \right\ra$ gives the average number of fermions in the region $B$. This expectation value can also be expressed in terms of the position density $\rho^{\psi}$ as follows:
\begin{eqnarray}
\left\la \psi(t) \right| \widehat{F}_d(B) \left| \psi(t) \right\ra 
&=& \int_B d^3 x \int \sum^n_{i=1} \delta({\bf x} - {\bf x}_i) \left\la \psi(t) \right| {\widehat P}(d^3x_1 \dots d^3x_n) \left| \psi(t) \right\ra \nonumber\\
&=& \int_B d^3 x \int_{{\mathbb R}^3}d^3x_1 \dots \int_{{\mathbb R}^3}d^3x_n  \sum^n_{i=1} \delta({\bf x} - {\bf x}_i) \rho^{\psi}({\bf x}_1, \dots,{\bf x}_n,t)\nonumber\\
&=& n \int_B d^3x_1 \int_{{\mathbb R}^3}d^3x_2 \dots \int_{{\mathbb R}^3}d^3x_n \rho^{\psi}({\bf x}_1, \dots,{\bf x}_n,t)\,,
\label{19.1}
\end{eqnarray}
where in the last equality we have used the fact that the density $\rho^{\psi}$ is completely symmetric in its arguments. 

\subsection{Pilot-wave model}\label{pwm}
We are now ready to present our pilot-wave model. As mentioned before we assume a finite fermion number. We will further elaborate on this issue in Section \ref{needregularization}. 

We introduce position beables for all the fermions, including the fermions in the Dirac sea. For the equilibrium density we take the density $\rho^{\psi}$ which corresponds to the position PVM ${\widehat P}(d^3x_1 \dots d^3x_n)$. As explained in Section \ref{general framework}, we can find guidance equations by considering the continuity equation for this density. First note that it follows from equation ({\ref{6}}) that the parts of the Hamiltonian that commute with the position POVM do not contribute to the guidance equation. Since in our case the bosonic Hamiltonian $\widehat{H}^B$ and interaction Hamiltonian $\widehat{H}_I$ commute with the position PVM ${\widehat P}(d^3x_1 \dots d^3x_n)$, we have that only $\widehat{H}^F_0$ contributes to the guidance equations. The fact that the bosonic Hamiltonian $\widehat{H}^B$ commutes with the position PVM is straightforward. The proof that the interaction Hamiltonian $\widehat{H}_I$ commutes with the position PVM is given in Appendix \ref{commutatorpositionpvm}.  

With the only contributions coming from the free fermionic Hamiltonian $\widehat{H}^F_0$, we find that the velocity field is given by
\begin{equation}
{\bf v}^{\psi}_k = \sum_{\lambda_1,\dots,\lambda_n,\xi} \sum_{^{a_1,\dots,a_n}_{a'_1,\dots,a'_n}}  \psi^{*}_{\lambda_1,a_1,\dots,\lambda_n,a_n,\xi } {\boldsymbol \alpha}^{(k)}_{a_1,\dots,a_n;a'_1,\dots,a'_n} \psi_{\lambda_1,a'_1,\dots,\lambda_n,a'_n,\xi} / \rho^{\psi}\,,
\label{21}
\end{equation}
where
\begin{equation}
{\boldsymbol \alpha}^{(k)}_{a_1,\dots,a_n;a'_1,\dots,a'_n} = \delta_{a_1a'_1} \dots {\boldsymbol \alpha}_{a_ka'_k} \dots \delta_{a_na'_n}\,.
\label{22}
\end{equation}
The guidance equations for the particles are now defined to be
\begin{equation}
\frac{{d} {\bf x}_k}{dt} = {\bf{v}}^{\psi}_k \,.
\label{23}
\end{equation}
These guidance equations are the natural generalizations of the guidance equations for the many particle Dirac equation \cite{bohm93}. The dynamics given by (\ref{21})--(\ref{23}) is also a special case of a dynamics proposed in \cite{durr02,durr031,durr032,tumulka03,durr04}: it is the `minimal process' associated with the position PVM ${\widehat P}(d^3x_1 \dots d^3x_n)$ and the Hamiltonian $\widehat{H}$, see e.g.\ Section 5 in \cite{durr04}.

The equilibrium distribution is given by $\rho^{\psi}$. Having an equilibrium distribution is a key requirement for showing that the pilot-wave model reproduces the predictions of standard quantum theory. We discuss this in more detail in Section \ref{quantum predictions} in the context of a regularized model.

Because the wavefunction $\psi_{\lambda_1,a_1,\dots,\lambda_n,a_n,\xi}({\bf x}_1,\dots,{\bf x}_n,t)$ is anti-symmetric under permutations of the label $1,\dots,n$ of the particle positions, also the particle dynamics, given by (\ref{21})--(\ref{23}), is invariant under such permutations. Therefore the particles are identical, i.e.\ they are not distinguished by any property such as mass, charge, flavour which could be associated to the label.
     
Since the particle label is actually redundant (it has no physical meaning), one could formulate a pilot-wave dynamics on the configuration space ${\mathbb R}^{3n}/S_n$, where $S_n$ is the group of permutations $\sigma$ of the set $\{1,\dots,n\}$ \cite{brownhr98}, instead of on the configurations space ${\mathbb R}^{3n}$. Alternatively, one could formulate a pilot-wave dynamics on the configuration space that is obtained from the space ${\mathbb R}^{3n}/S_n$ by removal of configurations for which two or more particle positions are the same \cite{brownhr98}. This latter configuration space can also be identified with the space ${}^n{\mathbb R}^{3}=\{S \subset{\mathbb R}^{3} | \# S=n \}$, which consists of $n$-element subsets of ${\mathbb R}^{3}$ \cite{goldstein042}. The dynamics on ${\mathbb R}^{3n}/S_n$ or ${}^n{\mathbb R}^{3}$ can be obtained from the dynamics on ${\mathbb R}^{3n}$ by a suitable projection.{\footnote{Interestingly, a dynamics on ${}^n{\mathbb R}^{3}$ can actually be formulated in general, i.e.\ also for quantum theories in which systems are not necessarily described by completely anti-symmetric wavefunctions \cite{goldstein042}.}}

In Bell's pilot-wave type model on the lattice, the beables are the fermion numbers at each lattice point \cite{bell872}. Hence, this approach also makes manifest that particles are considered to be identical. In our model, we can introduce a fermion number density
\begin{equation}
F_d({\bf x}) = \sum^n_{i=1}\delta({\bf x} - {\bf x}_i) 
\label{23.1}
\end{equation}
corresponding to the fermion positions ${\bf x}_1,\dots,{\bf x}_n$. The fermion number density is the natural generalization of Bell's fermion numbers to the continuum. In the rest of the paper, we interchangeably use the positions or the number density to characterize the fermions.{\footnote{Note that we attached several distinct meanings to the notion of particle. We have the quantum mechanical notion of a particle, whereby a particle is represented by its quantum state, and the notion of a particle in pilot-wave theory, where it is described by its position. The quantum mechanical notion of a particle can further be understood as in the standard representation of quantum field theory, which uses `particles' and `anti-particles', or as in the Dirac see representation. A similar ambiguity arises in our use of the notion of `fermion'. We hope it is clear from the context which notion we have in mind.}}

The fermion number density should not be confused with the fermionic (electric) charge density,{\footnote{The total charge density consists of the fermionic charge density plus the bosonic charge density.}} for which the operator is given by 
\begin{equation}
\widehat{Q}({\bf x})=\sum_{\lambda,a} q_\lambda \widehat{\psi}^\dagger_{\lambda,a}({\bf x})\widehat{\psi}_{\lambda,a}({\bf x})\,,
\label{23.4}
\end{equation}
where $q_\lambda$ is the charge of the particle with quantum numbers $\lambda$. The charge operator $\widehat{Q}=\int d^3x\widehat{Q}({\bf x})$ represents the sum of the charges of the positive and negative energy particles. The states $\left|{\bf x}_1,\lambda_1,a_1;\dots; {\bf x}_n,\lambda_n,a_n \right\ra$ are also eigenstates of the fermionic charge density operator, with eigenvalues $\sum^n_{i=1} q_{\lambda_i} \delta({\bf x} - {\bf x}_i)$. 

When Colin presented the Dirac sea pilot-wave model for quantum electrodynamics \cite{colin031,colin032,colin033}, only electrons and positrons were considered, in which case the fermion number density operator is proportional to the fermionic charge density operator. So in this particular case, one could equally well say that beables are introduced for charge density. In the model presented in this paper this can not be said; the beables correspond solely to fermion number density.

Fermion number superselection played a key role in the construction of the pilot-wave model. The pilot-wave model as such obtained now implies an even stronger notion of fermion number superselection, which is not available in standard quantum theory, and which is called {\em strong superselection}. This is explained in detail in \cite{colin06}. So note that if there was no superselection according to standard quantum theory, then our pilot-wave model would fail to reproduce the standard quantum predictions, simply because our model implies this even stronger notion of superselection.

\subsection{The need for regularization}\label{needregularization}
As mentioned before, the theory needs regularization. The regularization is required because of the usual field theoretical divergences. Further, in order to have well defined probability densities for the position beables, we need to deal with the non-existence of a generalization of the Lebesgue measure to an infinite dimensional space.

A possible way to regularize the pilot-wave model is as follows. We could assume that space is bounded and impose some suitable boundary conditions. For example we could assume that space is a cubic box of volume $V$ and that we have periodic boundary conditions for the field operators. For the Fourier expansions of the field operators $\widehat{\psi}_\lambda({\bf x})$ this means that we have to replace
\begin{equation}
 \int \frac{d^3 p}{\sqrt{(2\pi)^3}}  \to \sqrt{\frac{1}{V^3}} \sum_{\bf p}\,,
\label{27}
\end{equation}
where the possible momenta in the sum on the right hand side are given by ${\bf p}=(n_1,n_2,n_3)2\pi/L$, where $(n_1,n_2,n_3) \in {\mathbb Z}^3$ and $L$ is the length of a side of the cubic box. This makes the number of possible momenta countable. By further assuming an ultra-violet momentum cut-off we have that the number of possible momenta is finite. This implies that any possible state can only contain a finite number of particles. In particular, the filled Dirac sea will only contain a finite number of particles. Therefore also in the pilot-wave model we only need to introduce a finite number of position beables. We discuss this way of regularization in more detail in the following section. The introduction of this regularization will cause a slight modification of the equations presented in the previous section.

Apart from regularization problems, there would be another problem when we would take the naive limit of the number of fermions to infinity in the pilot-wave model presented in the previous subsection. The problem is the identification of an image of a familiar macroscopic world in the beable configuration \cite{tumulka06}. We would expect that we just have to count the number of fermions in different regions in physical space in order to find such an image. However, this procedure would not work for the following reason. We have seen that the quantity $\left\la \psi(t) \right| \widehat{F}_d(B) \left| \psi(t) \right\ra$ gives the average number of fermions in the region $B$. For states of physical interest, i.e.\ states which describe a finite number of positive energy particles and a finite number of holes in the negative energy sea, the average number of fermions in every region of non-zero measure, is infinite. For example, for the state $|0\ra$ we have (see Appendix \ref{expectationvaluevacuum} for the details of the calculation)
\begin{equation}
\left\la 0 \right| \widehat{F}_d(B) \left| 0 \right\ra = \frac{1}{(2\pi)^3} \sum_{\lambda,s} \int d^3 p \int_B d^3x = \infty \,.
\label{25}
\end{equation}
This means that every region in physical space looks the same if only the number of fermions is counted in those regions. As such, an image of the macroscopic world would clearly be absent. It is an open question whether there exists an alternative way of identifying such an image. The problem disappears when an ultra-violet momentum cut-off is introduced, since then
\begin{equation}
\left\la 0 \right| \widehat{F}_d(B) \left| 0 \right\ra = \frac{1}{(2\pi)^3} \sum_{\lambda,s} \int_{|{\bf p}| \le \Lambda} d^3 p \int_B d^3x =\frac{\Lambda^3}{3\pi^2} \sum_{\lambda} \int_{B} d^3x < \infty \,.
\label{28.1}
\end{equation}
We will discuss this in more detail in the following section.

\section{Ultra-violet momentum cut-off regularization}\label{ultravioletcutoff}
In this section, we regularize the theory by imposing an ultra-violet momentum cut-off $|{\bf p}| \le \Lambda$. As indicated in Section \ref{needregularization}, imposing the cut-off is actually not sufficient to completely regularize the pilot-wave model. We mentioned that a complete regularization could be obtained by assuming a bounded space and by assuming some suitable boundary conditions. We will not take this last step in regularizing the theory, since it would merely complicate some of the calculations, without changing the qualitative content. Nevertheless, at some places in this section we will implicitly assume a bounded space (e.g.\ in Section \ref{quantum predictions} where we discuss how the model reproduces the quantum predictions).

First, we will review the modifications to the equations that are caused by introducing the momentum cut-off. The most important difference is that the form of the guidance equations needs to be slightly modified in order to guarantee a pilot-wave dynamics which preserves the probability density $\rho^{\psi}$.

Then we will study what the typical beable configurations are corresponding to different macroscopic states. This is a prelude to the final subsection in which we study how our pilot-wave model reproduces the quantum predictions.

\subsection{Introducing the cut-off}
We introduce an ultra-violet momentum cut-off by making the replacement  $\int d^3 p \rightarrow \int_{|{\bf p}| \le \Lambda} d^3 p$ in the Fourier expansion of the field operators. This is equivalent to putting the creation and annihilation operators to zero for momenta higher than the cut-off. As a result the field anti-commutation relations change to
\begin{equation}
\{\widehat{\psi}_{\lambda,a}({\bf x}),\widehat{\psi}^\dagger_{\lambda',a'}({\bf x}')\}=\delta_{\lambda{\lambda'}}\delta_{aa'}\delta^{(\Lambda)}({\bf x}-{\bf x}')\,,
\label{c.1}
\end{equation}
where 
\begin{equation}
\delta^{(\Lambda)}({\bf x}) =\frac{1}{(2\pi)^3}   \int_{|{\bf p}| \le \Lambda} d^3p  e^{i{\bf p}\cdot {\bf x}}\,.
\label{c.2}
\end{equation}
Since the distribution $\delta^{(\Lambda)}({\bf x})$ does not vanish for ${\bf x} \neq {\bf 0}$, the field operators $\widehat{\psi}$  and $\widehat{\psi}^\dagger$ do not anti-commute anymore at different points in space. This causes some modifications in the equations of the pilot-wave theory, as we will soon see. 

First note that the distribution $\delta^{(\Lambda)}({\bf x})$ still acts as a $\delta$-distribution on the space of functions for which the Fourier series only contain momenta $|{\bf p}| \le \Lambda$, i.e.\ if 
\begin{equation}
f({\bf x})=  \frac{1 }{\sqrt{(2\pi)^3}} \int_{|{\bf p}| \le \Lambda}  d^3 p \widetilde{f}({\bf p}) e^{i{\bf p}\cdot {\bf x}}\,,
\label{c.3}
\end{equation}
then 
\begin{equation}
\int d^3x f({\bf x}) \delta^{(\Lambda)}({\bf x}-{\bf y})=f({\bf y})\,.
\label{c.4}
\end{equation}
On the other hand, if we consider two such functions $f$ and $g$ which have a momentum cut-off, then their product $fg$ generally has Fourier modes with momenta higher than the cut-off. This means that in general 
\begin{equation}
\int d^3x f({\bf x}) g({\bf x}) \delta^{(\Lambda)}({\bf x}-{\bf y})\neq f({\bf y}) g({\bf y})\,.
\label{c.5}
\end{equation}
This property will become important when we search for possible guidance equations.

By replacing the ordinary Dirac $\delta$-distribution by the modified distribution $\delta^{(\Lambda)}$, most of the equations of Section \ref{the model} are immediately generalized to the case in which we have a momentum cut-off. One can still introduce the states $\left|{\bf x}_1,\lambda_1,a_1;\dots; {\bf x}_n,\lambda_n,a_n \right\ra$ by application of the operators $\widehat{\psi}^\dagger$ onto the state $| 0_D \ra$, similarly as in (\ref{14.2}). However, now the states are normalized with respect to the distribution $\delta^{(\Lambda)}$, i.e.\
\begin{multline}
\la {\bf x}_1,\lambda_1,a_1;\dots; {\bf x}_n,\lambda_n,a_n \left|{\bf x}'_1,\lambda'_1,a'_1;\dots; {\bf x}'_n,\lambda'_n,a'_n \right\ra   \\
 =\frac{1}{n!}  \sum_\sigma (-1)^\sigma \delta^{(\Lambda)}({\bf x}_{\sigma(1)} -  {\bf x}'_1)\delta_{\lambda_{\sigma(1)}\lambda'_1}   \delta_{a_{\sigma(1)}a'_1}  \dots  \delta^{(\Lambda)}({\bf x}_{\sigma(n)} - {\bf x}'_n)\delta_{\lambda_{\sigma(n)}\lambda'_n}   \delta_{a_{\sigma(n)}a'_n}\,. 
\label{c.5001}
\end{multline}
Therefore different states are not orthogonal anymore (so that they form an overcomplete basis). As a result, the measure 
\begin{multline}
{\widehat P}(d^3x_1 \dots d^3x_n)= \sum_{\lambda_1,a_1,\dots,\lambda_n,a_n,\xi}  \left|{\bf x}_1,\lambda_1,a_1;\dots; {\bf x}_n,\lambda_n,a_n  \right\ra \otimes \left| \xi \right\ra \\
\times \left\la \xi \right| \otimes  \left\la{\bf x}_1,\lambda_1,a_1;\dots; {\bf x}_n,\lambda_n,a_n \right| d^3x_1 \dots d^3x_n 
\label{c.5002}
\end{multline}
is no longer projection valued but positive operator valued. 

Another consequence of the modified commutation relations of the field operators is that the states $\left|{\bf x}_1,\lambda_1,a_1;\dots; {\bf x}_n,\lambda_n,a_n \right\ra$ are no longer eigenstates of the fermion number density operator $\widehat{F}_d({\bf x})=\sum_{\lambda,a}\widehat{\psi}^\dagger_{\lambda,a}({\bf x})\widehat{\psi}_{\lambda,a}({\bf x})$, although they are still eigenstates of the fermion number operator $\widehat{F}_d$, with eigenvalue $n$.

Since the fermion number operator $\widehat{F}_d$ commutes with the Hamiltonian, the fermion number is conserved. Assuming fermion number superselection as before, we can assume a Hilbert space which contains only states with a fixed fermion number. In the Hilbert space with a fixed fermion number $n$, a quantum state $\left| \psi (t)\right\rangle$ can be expanded as
\begin{multline}
\left| \psi (t)\right\rangle =\sum_{\lambda_1,a_1,\dots,\lambda_n,a_n,\xi}  \int d^3x_1 \dots d^3x_n \\
\times \psi_{\lambda_1,a_1,\dots,\lambda_n,a_n,\xi}({\bf x}_1,\dots,{\bf x}_n,t) \left|{\bf x}_1,\lambda_1,a_1;\dots; {\bf x}_n,\lambda_n,a_n \right\ra \otimes \left| \xi \right\ra\,, 
\label{c.501}
\end{multline}
where the expansion coefficients $\psi_{\lambda_1,a_1,\dots,\lambda_n,a_n,\xi}({\bf x}_1,\dots,{\bf x}_n,t)$ are given by
\begin{equation}
\psi_{\lambda_1,a_1,\dots,\lambda_n,a_n,\xi}({\bf x}_1,\dots,{\bf x}_n,t) =\left(  \left\langle {\bf x}_1,\lambda_1,a_1;\dots; {\bf x}_n,\lambda_n,a_n \right| \otimes \left\langle \xi \right| \right) \left| \psi (t)\right\rangle \,,
\label{c.502}
\end{equation}
just as in (\ref{18}). From their definition it follows that the expansion coefficients are anti-symmetric under permutations of the indices $1,\dots,n$ and contain only Fourier modes with momenta smaller than the cut-off. The latter property can easily be seen to follow from the fact that the states $\left|{\bf x}_1,\lambda_1,a_1;\dots; {\bf x}_n,\lambda_n,a_n \right\ra$ only have Fourier modes with momenta smaller than the cut-off. This property will be important in evaluating integrals containing $\delta^{(\Lambda)}$-distributions.

To construct a pilot-wave model we take the measure ${\widehat P}(d^3x_1,\dots,d^3x_n)$ as the position POVM. For a state $| \psi \ra$ with a definite fermion number, the position density corresponding to the POVM ${\widehat P}(d^3x_1 \dots d^3x_n)$ is still of the form 
\begin{equation}
\rho^{\psi}({\bf x}_1, \dots,{\bf x}_n,t) = \sum_{\lambda_1,a_1,\dots,\lambda_n,a_n,\xi} \left| \psi_{\lambda_1,a_1,\dots,\lambda_n,a_n,\xi}({\bf x}_1,\dots,{\bf x}_n,t) \right|^2\,,
\label{c.503}
\end{equation}
where it is understood that the $\psi_{\lambda_1,a_1,\dots,\lambda_n,a_n,\xi}({\bf x}_1,\dots,{\bf x}_n,t)$ now only have Fourier modes with momenta smaller than the cut-off, but the continuity equation for the density gets an extra term, which is due to the fact that the commutator $\left[{\widehat P}(d^3x_1 \dots d^3x_n),\widehat{H}_I\right]$ is no longer zero. The fact that the commutator is no longer zero follows from the property (\ref{c.5}) above. 

In order to see what exactly happens, let us abandon the generality at this point and let us for simplicity just consider a single fermion. Similar expressions hold in the case of many fermions. For a one particle state the position density reads 
\begin{equation}
\rho^{\psi}({\bf x},t)=\sum_{\lambda ,a ,\xi} |\psi_{\lambda ,a ,\xi}({\bf x},t)|^2 \,.
\label{c.7}
\end{equation}
The position density satisfies the equation
\begin{equation}
\frac{\partial \rho^{\psi}({\bf x},t) }{\partial t} d^3 x +  \left\langle \psi(t) \right| i\left[  {\widehat P}(d^3 x) , {\widehat H}  \right]\left| \psi(t) \right\rangle=0\,,
\label{c.7.1}
\end{equation}
cf.\ Section \ref{general framework}, from which it follows that
\begin{equation}
\frac{\partial\rho^{\psi} }{\partial t} +  {\boldsymbol{\nabla}} \cdot \left({\bf v}^{\psi}  \rho^{\psi} \right) + g^{\psi} =0\,,
\label{c.8}
\end{equation}
where 
\begin{equation}
{\bf v}^{\psi} = \sum_{\lambda ,a,a' ,\xi} \psi^*_{\lambda ,a ,\xi}   {\boldsymbol{\alpha}}_{aa'} \psi_{\lambda ,a' ,\xi} / \rho^{\psi}
\label{c.9}
\end{equation}
is the usual velocity field and where
\begin{multline}
g^{\psi}({\bf x},t) = i \sum_{\lambda_1,\lambda_2,a_1,a_2,\xi_1,\xi_2} \int d^3y \big\la \xi_1 \big|\widehat{h}_{\lambda_1,a_1,\lambda_2,a_2}({\bf y})\big|\xi_2  \big\ra \delta^{(\Lambda)}({\bf x}-{\bf y}) \\
\times \left(  \psi^*_{\lambda_1 ,a_1 ,\xi_1}({\bf x},t) \psi_{\lambda_2 ,a_2 ,\xi_2}({\bf y},t)  -  \psi^*_{\lambda_1 ,a_1 ,\xi_1}({\bf y},t) \psi_{\lambda_2 ,a_2 ,\xi_2}({\bf x},t) \right) \,,
\label{c.10}
\end{multline}
is a contribution which comes from the term $\left\langle \psi(t) \right| i [  {\widehat P}(d^3 x) , {\widehat H}_I  ]\left| \psi(t) \right\rangle$ in equation (\ref{c.7.1}). This contribution does not vanish, since as mentioned above, the distribution $\delta^{(\Lambda)}({\bf x}-{\bf y})$ will not act as a $\delta$-distribution, because we integrate over a product of two functions whose Fourier expansion has a cut-off.{\footnote{Note in passing that $\int d^3 x g^{\psi}({\bf x},t) = 0$, so that the equation (\ref{c.8}) implies that, after integration over the complete space, $d (\int d^3 x \rho^{\psi}({\bf x},t))/ d t = 0$, which just expresses the conservation of fermion number.}}

The term $g^{\psi}$ implies that the field ${\bf v}^{\psi}$ can not serve as the velocity field for the beables anymore, since it would not preserve the density $\rho^{\psi}$. One possible way to introduce a dynamics which preserves the density $\rho^{\psi}$ is by supplementing the motion determined by the velocity field ${\bf v}^{\psi}$ by a jump process, as explained in \cite{durr02,durr031,durr032,tumulka03,durr04}. In this way the particle would move along a deterministic trajectory, determined by the velocity field ${\bf v}^{\psi}$, until a jump occurs. It is expected that the distance over which the particle jumps is typically of the order $1/\Lambda$.

Instead of introducing the jumps, which are stochastic, there probably also exist possibilities of keeping the dynamics deterministic. Suppose, for example, that $g^{\psi}$ is such that we can write
\begin{equation}
g^{\psi}({\bf x},t) = {\boldsymbol{\nabla}} \cdot \left(  {\boldsymbol{\nabla}} \nabla^{-2}  g^{\psi}({\bf x},t) \right) \,,
\label{c.11}
\end{equation}
where
\begin{equation}
\nabla^{-2}  g^{\psi}({\bf x},t) = - \int d^3 y \frac{g^{\psi}({\bf y},t)}{4\pi |{\bf x} - {\bf y}|}  \,.
\label{c.12}
\end{equation}
Actually checking that we can write $g^\psi$ as in $(\ref{c.11})$ would lead us too far, since $g^\psi$ depends on the bosonic operators and hence on the details of the quantization procedure. If we can write $g^\psi$ as in $(\ref{c.11})$ then the equation (\ref{c.8}) reduces to the continuity equation
\begin{equation}
\frac{\partial\rho^{\psi} }{\partial t} +  {\boldsymbol{\nabla}} \cdot \left(\left({\bf v}^{\psi} + {\widetilde {\bf v}}^{\psi}\right) \rho^{\psi} \right) =0\,,
\label{c.13}
\end{equation}
where
\begin{equation}
{\widetilde {\bf v}}^{\psi} =  {\boldsymbol{\nabla}} \left(\nabla^{-2} g^{\psi} \right) / \rho^{\psi}\,.
\label{c.14}
\end{equation}
Motivated by this continuity equation, we could introduce the guidance equation
\begin{equation}
\frac{{d} {\bf x}}{dt} = {\bf{v}^{\psi}} + {\widetilde {\bf v}}^{\psi}\,.
\label{c.15}
\end{equation}

Since the contribution $g^{\psi}$ disappears in the limit where the cut-off $\Lambda$ is taken to infinity, we suspect that the importance of the contribution $g^{\psi}$ for the dynamics of the beables decreases with an increasing cut-off $\Lambda$. Therefore we do not study the modified guidance equations in more detail.

With the introduction of a cut-off the equations of motion for both the quantum state and the beables become cut-off dependent. The next step would be to remove the cut-off by applying a suitable renormalization scheme and by taking the cut-off to infinity. This would require a rather careful analysis, which is beyond the scope of this paper. One of the problems associated with taking the cut-off to infinity is that the number of position beables would go to infinity again (assuming that we have a bounded space), so that the problem of defining a measure on an infinite dimensional space reappears. On the other hand, there could be a natural ultra-violet cut-off, so that taking the cut-off to infinity would not be required.

\subsection{Typical beable configurations for macroscopic states}\label{typicalbeableconfigurations}
In order for our pilot-wave model to reproduce the predictions of standard quantum theory it is crucial that different macroscopic states correspond to different typical beable configurations.  Having different typical beable configurations for different macroscopic situations namely implies that measurement results get recorded in the beable configuration. In the following section, we discuss the equivalence of our pilot-wave model with standard quantum theory in some detail. In this section, we first consider the question what the typical beable configuration is for a macroscopic state and whether such typical configurations exist in the first place. 

In order to be precise, we say that a property of a beable configuration is {\em typical} if the property holds for almost all beable configurations. With `almost all' we mean that the set of beable configurations which do not exhibit the property has very small $\rho^{\psi}({\bf x}_1,\dots,{\bf x}_n) d^3 x_1 \dots d^3 x_n$-measure, where $\rho^{\psi}$ is the position density. We say that a typical property holds for typical beable configurations.{\footnote{The language of typicality is borrowed from D\"urr {\em et al.}\ \cite{durr92} who use it in the context of justifying quantum equilibrium.}} 

For example, in non-relativistic pilot-wave theory a typical beable configuration for a macroscopic system is such that the actual position distribution in physical space gives, on some coarse-grained level, the classical image of the macroscopic object. For example, a typical beable configuration for the state of a table (given by the wavefunction of the table) gives the classical image of a table. 

In the pilot-wave model presented in this paper we also have particles for macroscopic objects, like tables, but here these particles move against a background of `vacuum particles' (of course, the distinction between particles belonging to the vacuum and the particles belonging to the system is only heuristic since the particles do not carry any label which would indicate such a distinction). As discussed in the previous section, this background of vacuum particles proves problematic if we do not introduce an ultra-violet momentum cut-off, since without cut-off the expected number of particles in any volume is infinite. It is only when we introduce a momentum cut-off that the expected number of particles becomes finite.  

Let us now study what a typical beable configuration will look like for a state describing a macroscopic system. For the vacuum state, we expect that the typical beable configuration is such that it gives an approximate uniform distribution of positions in physical space. By `approximate' we mean that the fluctuations in the distribution become negligible after suitable coarse-graining. For the state corresponding to a macroscopic system, we expect a distribution which is made up of a uniform distribution of particles corresponding to the vacuum and a distribution of particles corresponding to the system, where the latter is localized around the location of the macroscopic system. 

Due to technical complications we do not rigorously prove that our expectation is correct. Instead we provide arguments that support our expectation. Let us first consider the vacuum state. The physical vacuum is not given by the state $|0\ra$, which, in the language of standard quantum field theory, contains no particle or anti-particles, and which would be the vacuum if there were no interactions; due to the interactions the actual vacuum is much more complex. In the case there were only electromagnetic interactions, the physical vacuum could be thought of as some superposition of the state $|0\ra$ with states containing one particle--anti-particle pair, states containing two pairs, and so on. However, it is unclear to us what exactly the physical vacuum looks like if also the weak and strong interactions are brought into account. Therefore instead of considering the physical vacuum, we consider the state $|0\ra$ for simplicity. For the state $|0\ra$ the expected number of fermions in a volume $B$, which we denote by $n_0(B)$, is given by (see Appendix \ref{expectationvaluevacuum} for the details of the calculation)
\begin{equation}
n_0(B) = \left\la 0 \right| \widehat{F}_d(B) \left| 0 \right\ra = \frac{1}{(2\pi)^3} \sum_{\lambda,s} \int_{|{\bf p}| \le \Lambda} d^3 p \int_B d^3x =\frac{1}{8\pi^2} \Lambda^3 V(B)\,,
\label{28}
\end{equation}
where $V(B)$ is the volume of the region $B$. This means that the expected fermion number density is uniform. In order to find out whether this behaviour is typical for beable configurations corresponding to the state $\left| 0 \right\ra$, we calculate the standard deviation
\begin{equation}
\Delta_0 F_d(B) =  \sqrt{\left\la 0 \right| \widehat{F}_d(B)^2 \left|  0 \right\ra -   \left\la 0 \right| \widehat{F}_d(B) \left| 0 \right\ra^2  }=  \sqrt{\left\la 0 \right| \widehat{F}_d(B)^2 \left|  0 \right\ra -   n_0(B)^2  }\,.
\label{29}
\end{equation}
The standard deviation gives an estimate of the possible fluctuation of the fermion number in the region $B$. A small standard deviation would indicate that a uniform distribution is typical.

In order to calculate the standard deviation, we assume a spherical region with radius $b$, which we denote $B_s$. We consider radii $b$ which are of the order of the atomic distance $10^{-10}{\textrm{m}}=1{\textrm{\AA}}$ or larger. This means that $b$ is at least two orders of magnitude larger than the Compton wavelength of the electron, which is of the order $10^{-12}{\textrm{m}}$. Hence $b$ is also much bigger than the Compton wavelengths of the quarks and the other electrically charged leptons. At this point, we also assume the cut-off to be of the order of the Planck mass, i.e.\ $1.3 \cdot 10^{19}{\textrm{GeV}} \sim10^{35}/{\textrm{m}}$.

The details of this calculation are given in Appendix \ref{expectationvaluevacuum}. The result is that the fluctuation of the number of fermions in $B_s$ is 
\begin{equation}
\Delta_0 F_d(B_s) \approx 0.39 \sqrt{\Lambda} V(B_s)^{1/6}  \,.
\label{30.001}
\end{equation}
If the fluctuations in different regions were uncorrelated then according to classical statistical physics one would expect the fluctuation to be of the order of $n_0(B_s)^{1/2} \approx 0.11 \Lambda^{3/2}V(B_s)^{1/2}$, i.e.\ the square root of the number of particles in the region $B_s$ \cite[pp.\ 8-9]{landau80}. Since $\Delta_0 F_d(B_s) \approx 0.39 \sqrt{\Lambda} V(B_s)^{1/6}$ the fluctuation per unit volume decreases faster with increase of the volume than expected in the absence of correlations. So the fluctuations must be correlated.{\footnote{Similar conclusions were drawn by Heisenberg \cite{heisenberg34}. Heisenberg performed a calculation similar to ours. The main difference is that Heisenberg also introduced a coarse-graining over time (which significantly simplified the computations). He had the additional result that the fluctuation also decreases when there is a coarse-graining over larger time intervals.}} 

Since we do not consider the physical vacuum, but only the free vacuum $\left| 0 \right\ra$, the only cause for suppression of fluctuations comes from the mutual repulsion of the vacuum particles due to the Pauli exclusion principle. It is unclear what exactly would happen in the physical vacuum. If there were only electromagnetic interactions and only one type of charged particle then the fluctuations would be even more suppressed because of vacuum polarization effects. However, it is unclear what exactly happens if we include all types of particles and interactions.

Before giving a numerical estimate for the fluctuations, we first consider a state describing a macroscopic system. Let us assume a state of the form
\begin{multline}
\left| \varphi \right\ra = \frac{1}{\sqrt{m!}} \sum_{\lambda_1,s_1,\dots,\lambda_m,s_m} \int_{|{\bf p}_1| \le   \Lambda} d^3 p_1 \dots \int_{|{\bf p}_m| \le \Lambda} d^3 p_m  {\widetilde \varphi}_{\lambda_1,s_1,\dots,\lambda_m,s_m}({\bf p}_1,\dots,{\bf p}_m) \\
\times {\widehat c}^{\dagger}_{\lambda_1,s_1}({\bf p}_1) \dots  {\widehat c}^{\dagger}_{\lambda_m,s_m}({\bf p}_m)  \left| 0 \right\ra \,,
\label{30.0001}
\end{multline}
where the expansion coefficients ${\widetilde \varphi}_{\lambda_1,s_1,\dots,\lambda_m,s_m}({\bf p}_1,\dots,{\bf p}_m)$ are given by
\begin{equation}
{\widetilde \varphi}_{\lambda_1,s_1,\dots,\lambda_m,s_m}({\bf p}_1,\dots,{\bf p}_m)  = \frac{1}{\sqrt{m!}} \langle 0 | {\widehat c}_{\lambda_m,s_m}({\bf p}_m)  \dots   {\widehat c}_{\lambda_1,s_1}({\bf p}_1)   | \varphi \ra 
\label{30.0002}
\end{equation}
when all momenta are smaller than $\Lambda$ and by zero otherwise. Note that the state contains $m$ particles on top of the vacuum $\left| 0 \right\ra$ (not the vacuum $\left| 0_D \right\ra$) and no anti-particles. We will further assume that the state contains only `ordinary matter', i.e.\ up and down quarks and electrons (in fact we should also include the background of neutrinos, but as we will argue below, the particle number density of neutrinos is considerably smaller than the particle number density of the ordinary matter, so that this background can be ignored for our purposes).

Using the Fourier expansion
\begin{multline}
\begin{aligned}
&\varphi_{\lambda_1,a_1,\dots,\lambda_m,a_m}({\bf x}_1,\dots,{\bf x}_m)  \\
& = \sum_{s_1,\dots,s_m} \int_{|{\bf p}_1| \le \Lambda} \frac{d^3 p_1}{\sqrt{(2\pi)^3}} \dots \int_{|{\bf p}_m| \le \Lambda} \frac{d^3 p_m}{\sqrt{(2\pi)^3}} \sqrt{\frac{m_{\lambda_1}}{E_{\lambda_1}({\bf p}_1)}} u_{\lambda_1,s_1,a_1}({\bf p}_1)e^{i{\bf p}_1\cdot{\bf x}_1} \dots 
\end{aligned} \\
\sqrt{\frac{m_{\lambda_m}}{E_{\lambda_m}({\bf p}_m)}} u_{\lambda_m,s_m,a_m}({\bf p}_m)e^{i{\bf p}_m\cdot{\bf x}_m} {\widetilde \varphi}_{\lambda_1,s_1,\dots,\lambda_m,s_m}({\bf p}_1,\dots,{\bf p}_m) \,,
\label{30.0003}
\end{multline} 
the state can be written as
\begin{multline}
\left| \varphi \right\ra = \frac{1}{\sqrt{m!}} \sum_{\lambda_1,a_1,\dots,\lambda_m,a_m} \int d^3 x_1 \dots d^3 x_m \varphi_{\lambda_1,a_1,\dots,\lambda_m,a_m}({\bf x}_1,\dots,{\bf x}_m)\\
\times {\widehat \psi}^{\dagger}_{\lambda_1,a_1}({\bf x}_1) \dots  {\widehat \psi}^{\dagger}_{\lambda_m,a_m}({\bf x}_m)  \left| 0 \right\ra \,,
\label{30.0004}
\end{multline}
where
\begin{equation}
\varphi_{\lambda_1,a_1,\dots,\lambda_m,a_m}({\bf x}_1,\dots,{\bf x}_m) = \frac{1}{\sqrt{m!}} \langle 0 | {\widehat \psi}_{\lambda_m,a_m}({\bf x}_m)  \dots  {\widehat \psi}_{\lambda_1,a_1}({\bf x}_1) | \varphi \ra \,.
\label{30.0005}
\end{equation}
Assuming the state $\left| \varphi \right\ra $ to be normalized to 1, we have 
\begin{eqnarray}
1&=& \sum_{\lambda_1,a_1,\dots,\lambda_m,a_m} \int d^3 x_1 \dots d^3 x_m |\varphi_{\lambda_1,a_1,\dots,\lambda_m,a_m}({\bf x}_1,\dots,{\bf x}_m)|^2 \nonumber\\
&=& \sum_{\lambda_1,s_1,\dots,\lambda_m,s_m} \int_{|{\bf p}_1| \le \Lambda} d^3 p_1 \dots \int_{|{\bf p}_m| \le \Lambda} d^3 p_m |{\widetilde \varphi}_{\lambda_1,s_1,\dots,\lambda_m,s_m}({\bf p}_1,\dots,{\bf p}_m)|^2\,.
\label{30.0006}
\end{eqnarray}

As with the vacuum state, our representation of a macroscopic system by the state $\left| \vp \right\ra $ is not entirely accurate. For the state $\left| \vp \right\ra$, the only interaction between the macroscopic system and the vacuum is caused by the Pauli exclusion principle. In general, we also have to include the standard interactions between the system and the vacuum, like the electromagnetic interaction, which will cause a reorganization of the vacuum particles. Again we only know what would happen if we only had electromagnetic interaction. In that case, we would have, in the standard particle--anti-particle picture, the appearance of virtual particle--anti-particle pairs around electrically charged particles. 

We want to calculate the expectation value and standard deviation for the fermion number for a region $B$ of the order of the atomic distance $10^{-10}{\textrm{m}}=1{\textrm{\AA}}$ or larger, just as we did for the vacuum $|0\ra$. In order to simplify the analysis, we assume that the state $\left| \varphi \right\ra$ represents particles that are approximately localized within this region $B$. This means that most of the support of the wavefunction $\varphi_{\lambda_1,a_1,\dots,\lambda_m,a_m}({\bf x}_1,\dots,{\bf x}_m)$ is within the region $B^m \subset {\mathbb R}^{3m}$. Note that the wavefunction cannot have all of its support within the region $B^m$, because it is a superposition of positive energy states (i.e.\ a superposition of the functions $u({\bf p})$). At best, most of the support can be within a region of the order of the Compton wavelength of the lightest particle under consideration, which in this case is the electron Compton wavelength $10^{-12}$m (we are assuming that the state represents ordinary matter). The effect of the cut-off on the localizability of the wavefunction is hereby negligible, since it only becomes important on scales of the order of $1/\Lambda$, which is the Planck length when $\Lambda$ is the Planck mass. Since we are interested in regions $B$ with a radius of the order of the atomic distance $10^{-10}{\textrm{m}}=1{\textrm{\AA}}$ or larger, the fact that most of the support can at best be within a region of the order of the electron Compton wavelength, will only be of minor importance.

The details of the calculation are given in Appendix \ref{expectationvaluemacroscopic}. We have that the expected number of particles in the region $B$ is given by 
\begin{equation}
\left\la \varphi \right| \widehat{F}_d(B) \left| \varphi \right\ra \approx  m +n_0(B) \,.
\label{30.1}
\end{equation}
For a region ${\tilde B} \subset {\mathbb R}^{3} \setminus B$ we have that  
\begin{equation}
\left\la \varphi \right| \widehat{F}_d({\tilde B}) \left| \varphi \right\ra \approx n_0({\tilde B})\,.
\label{30.03}
\end{equation}
So in the region $B$ the expected number of particles is approximately given by $m$, the number of particles corresponding to the macroscopic object, plus $n_0(B)$, the number of vacuum particles in the region $B$. In a region ${\tilde B}$ outside $B$ the expected number of particles is given by the number of vacuum particles in that region. 

The standard deviation is given by
\begin{equation}
\Delta_\varphi F_d(B) =  \sqrt{\left\la \varphi \right| \widehat{F}_d(B)^2 \left| \varphi \right\ra - \left\la \varphi \right| \widehat{F}_d(B) \left| \varphi \right\ra^2 } \approx \Delta_0 F_d(B) \,.
\label{31}
\end{equation}
For a region ${\tilde B}$ outside $B$, we have the same expression for the standard deviation. 

Let us now consider the question whether states corresponding to macroscopically distinct systems will correspond to different typical beable configurations. Two states, say $|\phi_1\ra$ and $|\phi_2\ra$, will correspond to distinct beable configurations if there exists a region $B$ for which
\begin{equation}
|\left\la \varphi_1 \right| \widehat{F}_d(B) \left| \varphi_1 \right\ra - \left\la \varphi_2 \right| \widehat{F}_d(B) \left| \varphi_2 \right\ra| \gg {\textrm{max}}(\Delta_{\varphi_1} F_d(B),\Delta_{\varphi_2} F_d(B)) \approx  \Delta_0 F_d(B) \,.
\label{31.001}
\end{equation}
In order to address the question let us consider a specific example. Consider two states, one which represents solid matter, say graphite, in a spherical region $B_s$ with radius $b$, and one which represents, say, air in that region. What should now be the volume $V(B_s)$ of the region $B_s$ in order for the states to correspond to different typical fermion number configurations in $B_s$? Answering this question will give us an idea whether macroscopically distinct systems will correspond to different typical beable configurations. If we find that we only need a small volume $V(B_s)$, e.g.\ microscopically small, then this would suggest that the answer to the latter question is affirmative. Now, since the state of the solid matter has a considerably higher fermion number density in $B_s$ than the state representing air, the volume $V(B_s)$ should be such that 
\begin{equation}
m \gg  \Delta_0 F_d(B_s)\,,
\label{31.002}
\end{equation}
where $m$ is the number of fermions corresponding to the solid matter contained in $B_s$. The fermion number density of graphite approximately given by $4.2 \cdot 10^{30}/{\textrm{m}}^3$ (1 carbon atom consists of 42 fermions, namely 18 up quarks, 18 down quarks and 6 electrons; $0.012$kg carbon contains $6\cdot 10^{23}$ atoms and the mass density of graphite is approximately given by $2\cdot 10^{3}{\textrm{kg}}/{\textrm{m}}^3$), so that $m \approx V(B_s) 4.2 \cdot 10^{30}/{\textrm{m}}^3$. We hereby ignored the background of neutrinos, since the density of neutrino is considerably smaller than the fermion density of graphite.{\footnote{The neutrino density on earth is believed to be dominated by relic neutrinos, i.e.\ the neutrinos which decoupled within the first 10 seconds after the Big Bang. Cosmological considerations lead to an estimate of a neutrino density of approximately $3.4\cdot 10^{8}/{\textrm{m}}^3$ \cite[p.\ 222]{klapdor-kleingrothaus00}. For comparison, the sun, which is the main steady source of neutrinos for the earth, gives a density of approximately $3.3 \cdot 10^{6}/{\textrm{m}}^3$ \cite[p.\ 14]{bahcall89}. The rest of the stars of our galaxy together only yield a density which is at most $10^{-8}$ times the density of solar neutrinos \cite[pp.\ 165-166]{bahcall89}. Other sources of neutrinos are supernovae. The well-known supernova SN1987A in the Large Magellanic Cloud, a nearby dwarf galaxy, is estimated to have produced a density of at most $10/{\textrm{m}}^3$, and this over a period of less than 15 seconds \cite[pp.\ 443]{bahcall89}. A typical supernova in our galaxy is estimated to produce a density which is 25 times larger \cite[pp.\ 437]{bahcall89}.}} Taking the cut-off $\Lambda$ of the order of the Planck mass $10^{35}/{\textrm{m}}$, we have that \eqref{31.002} is valid when the radius $b$ of $B_s$ satisfies $b \gg 2.6 \cdot 10^{-6}{\textrm{m}}$. This is a rather large lower bound (for comparison the diameter of a human hair typically ranges from $10^{-4}{\textrm{m}}$ to $10^{-5}{\textrm{m}}$). However as mentioned before, this result would be modified when all the interactions would be taken into account. If there were only electromagnetic interactions and one type of particle then we know that the result would improve. 

Note also that the estimate is cut-off dependent. A lower cut-off would lead to a decrease of the lower bound. A higher cut-off on the other hand would lead to an increase of the lower bound. This seems to imply that taking the cut-off to infinity in a renormalization process would obscure the identification of macroscopic objects in the beable configuration. On the other hand, in the case there is a natural cut-off, which is not too high, there is no problem with such an identification.

\subsection{Reproducing the quantum predictions}\label{quantum predictions}
Given that macroscopically distinct states imply distinct typical beable configurations, it is straightforward to show that our pilot-wave model reproduces the predictions of standard quantum field theory. The arguments are similar to the ones used to show that non-relativistic pilot-wave theory reproduces the predictions of standard quantum mechanics. For definiteness we repeat these arguments here. We closely follow the original exposition of Bohm, who was the first to present a convincing analysis for non-relativistic pilot-wave theory \cite{bohm521} .

Since quantum theory makes probabilistic statements about outcomes of measurements, we just consider the measurement situation in detail. According to standard quantum theory, or at least according to some versions of it, a measurement situation can be described as follows. The wavefunction $\psi$ (which is the expansion coefficient as defined in equation ({\ref{c.501}})), which describes the system under observation together with the measurement device, macroscopic pointer and environment, evolves into a superposition $\sum_i c_i \psi_i$, where the $\psi_i$ are normalized to 1. The different $\psi_i$ correspond to the different possible outcomes of the measurements, i.e.\ the different $\psi_i$ describe the different possible macroscopic pointer configurations which indicate the different possible measurement outcomes, like for example a needle pointing in different directions. The wavefunction then collapses to one of the $\psi_i$ in the superposition, say $\psi_k$, and this with probability $|c_k|^2$. The state $\psi_k$ represents one of the possible macroscopic pointer configurations and hence also the corresponding outcome of the measurement. So the collapse ensures that measurements have definite outcomes, whereby the outcome is determined by the wavefunction through the eigenstate-eigenvalue link. Of course, as is well known, it is rather ambiguous when the collapse exactly occurs and this ambiguity forms the core of the measurement problem.

In our pilot-wave model the wavefunction evolves according to the Schr\"o\-din\-ger equation at all times. The wavefunction never undergoes collapse. Outcomes of measurements are not recorded in the wavefunction, like in standard quantum theory, but in the beable configuration. This is seen as follows. In a measurement situation the wavefunction evolves into a superposition $\sum_i c_i \psi_i$, just as in standard quantum theory, but there is no subsequent collapse. Since the $\psi_i$ correspond to different macroscopic situations, like for example a needle pointing in different directions, they will have different typical beable configurations. This implies that the $\psi_i$ have minimal overlap, i.e.\ that they have approximately disjoint supports in configuration space. This means that the actual beable configuration $({\bf x}_1,\dots,{\bf x}_n)$ will be in the support of only one of the wavefunctions $\psi_i$, say $\psi_k$, or in other words the actual beable configuration will only be a typical configuration for one of the wavefunctions $\psi_i$. In this way, the beable configuration displays the measurement outcome. Since in equilibrium, the probability for the beable configuration $({\bf x}_1,\dots,{\bf x}_n)$ to be in the support of $\psi_k$ is given by $|c_k|^2$, we recover the quantum probabilities. 

Further, if the different wavefunctions $\psi_i$ stay approximately non-over\-lap\-ping at later times, then one can easily check that only the wavefunction $\psi_k$ determines the velocity field for the beable configuration. Since the other packets $\psi_i$, with $i \neq k$, play no role in determining the velocity field anymore, one can simply remove them from the description of the beable configuration. This is called an {\em effective collapse} and clearly explains the success of the ordinary collapse in standard quantum theory.

\section{Note on the predicted violation of fermion number conservation in the Standard Model}\label{violationfermionnumber}
The assumption that fermion number $F$ is conserved is a key ingredient that allows us to construct our pilot-wave model. This assumption is justified in the energy regime where the electroweak gauge symmetry is broken. In the same regime, one actually has that both quark number $Q$ and lepton number $L$ are conserved. The quark number is defined as the number of quarks minus the number of anti-quarks $Q=n_q - n_{\bar q}$. The lepton number $L=n_l - n_{\bar l}$ is the number of leptons minus the number of anti-leptons. The fermion number $F$ is the sum of the two, i.e.\ $F=Q+L$. Conservation of both quark number $Q$ and lepton number $L$ actually allows us to also introduce additional beables corresponding to the particle type, i.e.\ quark or lepton. In such a pilot-wave model the actual particles would not only have a position but also a label indicating whether it belongs to the family of quarks or the family of leptons.

However, the Standard Model predicts a violation of conservation of all three numbers $F$, $Q$ and $L$ for sufficiently high energies. The violations of conservation of these quantum numbers are so-called anomalies.{\footnote{A clear introduction to anomalies is given in \cite{treiman85}, \cite[pp.\ 1-102]{jackiw95} and \cite[pp.\ 217-235]{greiner02}.}} This means that classically, i.e.\ before quantization, these quantum numbers are conserved. It is only after quantization that they are not conserved anymore. Before quantization the theory is invariant under the global $U(1)$ symmetries $\psi_\lambda \to e^{i\alpha} \psi_\lambda$, where the index $\lambda$ either ranges over all possible values, or only the values corresponding to quarks, or only the values correspond to leptons. By Noether's theorem this respectively gives us conservation of $F$, $Q$ and $L$. When quantizing one needs to give up either these global symmetries or the electroweak gauge $SU(2) \times U(1)_Y$ symmetry. More specifically, it is the regularization that goes with the quantization that requires that either of these symmetries is given up. Since the gauge invariance is important to show renormalizability, the global $U(1)$ symmetries are given up, so that the quantum numbers $F$, $Q$ and $L$ are no longer conserved.{\footnote{For example, if we were to apply a similar cut-off regularization as in Section \ref{ultravioletcutoff} to the `full-blown' electroweak theory (i.e.\ not merely the spontaneously broken sector of the electroweak theory), then the fermion number could still be conserved since this type of regularization breaks the electroweak gauge symmetry.}} The possibility of non-conservation of these quantum numbers is important in cosmology since it might yield an explanation for baryogenesis, see e.g.\ \cite{dolgov92,dolgov97,cline06}. 

Interestingly, often a heuristic derivation of anomalies is given in the Dirac sea picture (see e.g.\ Jackiw \cite[pp.\ 333-337]{treiman85}, \cite[pp.\ 22-26]{jackiw95}, \cite{jackiw99}, and Greiner {\em et al.}\ \cite[pp.\ 219-221]{greiner02}). Maybe this heuristic derivation sheds a light on how to construct a pilot-wave model in the case when fermion number conservation is violated.  

An alternative would be to consider the quantum number $B-L$ where $B$ is the baryon number $B=Q/3$. This quantum number is conserved in the standard model. It is also conserved in unification theories like e.g.\ the $SU(5)$ unification theory \cite{mohapatra02}. Theories that conserve $B-L$ also go by the name $B-L$ theories \cite{mohapatra02}. Maybe one can introduce beables by exploiting the conservation of the quantum number $B-L$.

\section{Outlook}
We further developed the Dirac sea pilot-wave model originally presented by Colin and Bohm {\em et al.} Several points need further attention. First of all, the model relies on the conservation of the fermion number $F$. Since the Standard Model predicts a violation of fermion number conservation for sufficiently high energies, our pilot-wave model should be further developed in order to be able to deal with this possibility. It is unclear how this could be done in a way that preserves determinism. An alternative would be to try to introduce beables for another quantity which is conserved according to the Standard Model, like baryon number minus lepton number $B-L$. 

We have further shown how the pilot-wave model can be regularized by introducing an ultra-violet momentum cut-off and by assuming a finite space. In this context we also discussed the emergence of the familiar macroscopic world in the model. Although we gave arguments to support our expectation that the familiar macroscopic world does emerge, a more conclusive analysis would be desired.

In our model no beables were introduced for the bosonic degrees of freedom of the quantum state. If the beables in our model provide an image of the familiar classical world, as we expect, then there is actually no need to introduce beables also for the bosonic quantum fields. Bell, for example, did not introduce beables for the bosonic degrees of freedom in his lattice model. On the other hand, one could introduce beables for the bosonic degrees of freedom by using the field beable approach initiated by Bohm for the quantized electromagnetic field. Another possibility would be to introduce particle beables also for the bosons, although it is unclear for the moment how this could be done in a satisfactory way. In connection to this, it might also be interesting to look at the supersymmetric version of the Standard Model. Since the supersymmetry implies a symmetry between bosons and fermions, maybe an ontology which treats bosons and fermions in a similar way suggests itself in this context. 

\section*{Acknowledgements}
Discussions with Detlef D\"urr and the Munich group, Thomas Durt, Sheldon Goldstein, Justin Khoury, Constantinos Skordis, Roderich Tumulka and Hans Westman are gratefully acknowledged. Special thanks also to Sheldon Goldstein and Roderich Tumulka for detailed comments on the draft. S.\ Colin is further grateful to Thomas Durt for support and guidance during his Ph.D., during which the ideas for this work developed. Research at Perimeter Institute is supported in part by the Government of Canada through NSERC and by the Province of Ontario through MEDT. 

\appendix
\section{Commutator of the position PVM and the interaction Hamiltonian}\label{commutatorpositionpvm}
We show that ${\widehat P}(d^3x_1 \dots d^3x_n)$ commutes with $\widehat{H}_I$, where ${\widehat P}(d^3x_1 \dots d^3x_n)$ and $\widehat{H}_I$ are respectively given in (\ref{17}) and (\ref{10}) (no cut-off is introduced here). 

In order to do so, first note that if an operator ${\widehat A}$ commutes with $\widehat{H}_I$ and with ${\widehat h}_{\lambda,a,\lambda',a'}({\bf x})$ for all values of $\lambda,\lambda',a,a',{\bf x}$, then $\sum_{\lambda,a}{\widehat \psi}^\dagger_{\lambda,a}({\bf x})  {\widehat A} {\widehat \psi}_{\lambda,a}({\bf x})$ commutes with $\widehat{H}_I$. This follows easily by using the commutation relations
\begin{eqnarray}
\left[ {\widehat \psi}_{\lambda,a}({\bf x}),\widehat{H}_I\right] &=& \sum_{\lambda',a'}{\widehat h}_{\lambda,a,\lambda',a'}({\bf x}){\widehat \psi}_{\lambda',a'}({\bf x}) \,, \nonumber\\ 
\left[ {\widehat \psi}^\dagger_{\lambda,a}({\bf x}),\widehat{H}_I\right] &=& - \sum_{\lambda',a'}{\widehat \psi}^\dagger_{\lambda',a'}({\bf x})  {\widehat h}_{\lambda',a',\lambda,a}({\bf x})\,.
\label{a0.1}
\end{eqnarray}

If we now take ${\widehat A} = \sum_{\xi} \left| 0_D \right\ra \otimes \left| \xi \right\ra \left\la \xi \right| \otimes  \left\la 0_D \right|$, then ${\widehat A}$ commutes with $\widehat{H}_I$ and with ${\widehat h}_{\lambda,a,\lambda',a'}({\bf x})$ for all values of $\lambda,\lambda',a,a',{\bf x}$ so that $\sum_{\lambda,a}{\widehat \psi}^\dagger_{\lambda,a}({\bf x}_1)  {\widehat A} {\widehat \psi}_{\lambda,a}({\bf x}_1) d^3x_1= P(d^3x_1)$ commutes with $\widehat{H}_I$. Then by induction and the fact that 
\begin{equation}
{\widehat P}(d^3x_1 \dots d^3x_n) = \frac{1}{n} \sum_{\lambda,a}{\widehat \psi}^\dagger_{\lambda,a}({\bf x}_n) {\widehat P}(d^3x_1 \dots d^3x_{n-1}) {\widehat \psi}_{\lambda,a}({\bf x}_n) d^3x_n\,,
\label{a0.2}
\end{equation}
we find that ${\widehat P}(d^3x_1 \dots d^3x_n)$ commutes with $\widehat{H}_I$.

As we explain in Section \ref{ultravioletcutoff}, this property no longer holds if a momentum cut-off is introduced. The reason is that (\ref{a0.1}) no longer holds in that case.

\section{Expectation value and standard deviation of the fermion number for the vacuum $|0\ra$}\label{expectationvaluevacuum}
In this appendix, we calculate the expectation value
\begin{equation}
n_0(B)=\langle 0|\widehat{F}_d(B)|0\rangle =\int_{B} d^3x \langle 0|\widehat{F}_d({\bf x})|0\rangle 
\label{a1}
\end{equation}
and the standard deviation 
\begin{equation}
\Delta_0 F_d(B)=\sqrt{\langle 0|\widehat{F}^2_d(B)|0\rangle - n_0(B)^2}
\label{a2}
\end{equation} 
for the fermion number in a region $B$ for the vacuum $|0\rangle$. In the calculations we use an ultra-violet momentum cut-off $|{\bf p}|\le \Lambda$. The cut-off is assumed to be at the Planck scale. 

The expectation value for the fermion number in the region $B$ is given by
\begin{eqnarray}
n_0(B)&=& \sum_{\lambda,a}\int_{B} d^3x \langle 0|{\widehat \psi}^\dagger_{\lambda,a}({\bf x}){\widehat \psi}_{\lambda,a}({\bf x})|0\rangle \nonumber\\
&=&\frac{1}{(2\pi)^3} \sum_{\lambda,a,s_1,s_2} \int_{B} d^3x \int_{|{\bf p}_1|\le \Lambda} d^3 p_1 
\int_{|{\bf p}_2|\le \Lambda} d^3 p_2\frac{m_\lambda}{\sqrt{E_\lambda({\bf p}_1)E_\lambda({\bf p}_2})}\nonumber\\
&& \quad \langle 0|\widehat{d}_{\lambda,s_1}({\bf p}_1)\widehat{d}^\dagger_{\lambda,s_2}({\bf p}_2)|0\rangle v^\dagger_{\lambda,s_1,a}({\bf p}_1)v_{\lambda,s_2,a}({\bf p}_2) e^{i({\bf p}_1-{\bf p}_2)\cdot{\bf x}}  \nonumber\\ 
&=& \frac{1}{(2\pi)^3} \sum_{\lambda,s}\int_{|{\bf p}|\le \Lambda} d^3 p \int_{B} d^3x\nonumber\\ 
&=& \frac{\Lambda^3}{3\pi^2} \sum_{\lambda} \int_{B} d^3x\,,\nonumber\\ 
&=& \frac{8 }{\pi^2} \Lambda^3 V(B) \,,
\label{a3}
\end{eqnarray}
where we have used the normalization of the spinors 
\begin{equation}
\sum_av^{\dagger}_{\lambda,s_1,a}({\bf p})v_{\lambda,s_2,a}({\bf p})= \delta_{s_1s_2} E_\lambda({\bf p})/m_\lambda
\label{a3.1}
\end{equation} 
in the third equality. In the last equality we used the fact that $\sum_{\lambda}1=24$ (we have six types of quarks, which can each have three different colours, and six types of leptons). We also introduced the notation $V(B)$ for the volume of $B$. Since the expectation value is proportional to the volume of the region $B$, the expected fermion number density is uniform. 

In order to get the standard deviation, we first calculate
\begin{equation}
\langle 0|\widehat{F}^2_d(B)|0\rangle=\sum_{\lambda,\lambda',a,a'} \int_{B} d^3x \int_{B}d^3y \langle 0|{\widehat \psi}^\dagger_{\lambda,a}({\bf x}){\widehat \psi}_{\lambda,a}({\bf x}){\widehat \psi}^\dagger_{\lambda',a'}({\bf y}){\widehat \psi}_{\lambda',a'}({\bf y})|0\rangle \,.
\label{a4}
\end{equation}
Since any field operator ${\widehat \psi}$ is a superposition of operators $\widehat{c}$ and $\widehat{d}^\dagger$, there will be two non-zero contributions to this expression: one coming from the product $\widehat{d}\widehat{d}^\dagger\widehat{d}\widehat{d}^\dagger$ and one coming from the product $\widehat{d}\widehat{c}\widehat{c}^\dagger\widehat{d}^\dagger$. We call these contributions respectively $C_I$ and $C_{II}$, so that $\langle 0|\widehat{F}^2_d(B)|0\rangle=C_{I}+C_{II}$. We have that
\begin{multline}
C_{I}=\frac{1}{(2\pi)^6}\sum_{\lambda,\lambda',a,a',s_k} \int_{B} d^3x \int_{B}d^3y  \int_{|{\bf p}_k|\le \Lambda}d^3 p_1 d^3 p_2 d^3 p_3 d^3 p_4 \\
\frac{m_\lambda m_{\lambda'}}{\sqrt{E_\lambda({\bf p}_1)E_\lambda({\bf p}_2)E_{\lambda'}({\bf p}_3)E_{\lambda'}({\bf p}_4)}} \langle 0| {\widehat d}_{\lambda,s_1}({\bf p}_1){\widehat d}^\dagger_{\lambda,s_2}({\bf p}_2){\widehat d}_{\lambda',s_3}({\bf p}_3){\widehat d}^\dagger_{\lambda',s_4}({\bf p}_4)|0\rangle \\
\times v^\dagger_{\lambda,s_1,a}({\bf p}_1)v_{\lambda,s_2,a}({\bf p}_2)v^\dagger_{\lambda',s_3,a'}({\bf p}_3)v_{\lambda',s_4,a'}({\bf p}_4) e^{i({\bf p}_1-{\bf p}_2)\cdot{\bf x}} e^{i({\bf p}_3-{\bf p}_4)\cdot{\bf y}}\,.
\label{a5}
\end{multline}
Using the canonical anti-commutation relations ({\ref{11.1}}), we find  
\begin{equation}
 \langle 0|d_{\lambda,s_1}({\bf p}_1)d^\dagger_{\lambda,s_2}({\bf p}_2)d_{\lambda',s_3}({\bf p}_3)d^\dagger_{\lambda',s_4}({\bf p}_4)|0\rangle = \delta_{s_1s_2} \delta_{s_3s_4}\delta({\bf p}_1-{\bf p}_2) \delta({\bf p}_3-{\bf p}_4)\,,
\label{a6}
\end{equation} 
so that after integrating over ${\bf p}_2$ and ${\bf p}_4$, we get 
\begin{equation}
C_{I}=\frac{1}{(2\pi)^6}\sum_{\lambda,\lambda',s_1,s_3}  \int_{|{\bf p}_1|\le \Lambda}d^3 p_1\int_{|{\bf p}_3|\le \Lambda} d^3 p_3 \int_{B} d^3x \int_{B}d^3y = n_0(B)^2\,.
\label{a7}
\end{equation}
Now we compute $C_{II}$. We find that 
\begin{multline}
C_{II}=\frac{1}{(2\pi)^6}\sum_{\lambda,\lambda',a,a',s_k} \int_{B} d^3x \int_{B}d^3y  \int_{|{\bf p}_k|\le \Lambda}d^3 p_1 d^3 p_2 d^3 p_3 d^3 p_4 \\
\frac{m_\lambda m_{\lambda'}}{\sqrt{E_\lambda({\bf p}_1)E_\lambda({\bf p}_2)E_{\lambda'}({\bf p}_3)E_{\lambda'}({\bf p}_4)}} \langle 0|\widehat{d}_{\lambda,s_1}({\bf p}_1)\widehat{c}_{\lambda,s_2}({\bf p}_2)\widehat{c}^\dagger_{\lambda',s_3}({\bf p}_3)\widehat{d}^\dagger_{\lambda',s_4}({\bf p}_4)|0\rangle \\
\times v^\dagger_{\lambda,s_1,a}({\bf p}_1)u_{\lambda,s_2,a}({\bf p}_2)u^\dagger_{\lambda',s_3,a'}({\bf p}_3)v_{\lambda',s_4,a'}({\bf p}_4) e^{i({\bf p}_1+{\bf p}_2)\cdot{\bf x}} e^{-i({\bf p}_3+{\bf p}_4)\cdot{\bf y}}\,.
\label{a8}
\end{multline}
Using
\begin{equation}
\langle 0|\widehat{d}_{\lambda,s_1}({\bf p}_1)\widehat{c}_{\lambda,s_2}({\bf p}_2)\widehat{c}^\dagger_{\lambda',s_3}({\bf p}_3)\widehat{d}^\dagger_{\lambda',s_4}({\bf p}_4)|0\rangle =  \delta_{\lambda \lambda'}\delta_{s_1s_4}\delta_{s_2s_3}\delta({\bf p}_1-{\bf p}_4) \delta({\bf p}_2-{\bf p}_3)
\label{a9}
\end{equation}
and integrating over ${\bf p}_3$ and ${\bf p}_4$, we get 
\begin{eqnarray}
C_{II}&=& \frac{1}{(2\pi)^6} \sum_{\lambda,a,a',s_1,s_2} \int_{B} d^3x \int_{B}d^3y \int_{|{\bf p}_1|\le \Lambda}d^3 p_1 \int_{|{\bf p}_2|\le \Lambda}d^3 p_2 \frac{m^2_\lambda}{E_\lambda({\bf p}_1) E_\lambda({\bf p}_2)} \nonumber\\
&& \quad v^\dagger_{\lambda,s_1,a}({\bf p}_1)u_{\lambda,s_2,a}({\bf p}_2)u^\dagger_{\lambda,s_2,a'}({\bf p}_2)v_{\lambda,s_1,a'}({\bf p}_1) e^{i({\bf p}_1+{\bf p}_2)\cdot({\bf x}-{\bf y})} \nonumber\\
&=& \frac{1}{(2\pi)^6} \sum_{\lambda} \int_{B} d^3x \int_{B}d^3y \int_{|{\bf p}_1|\le \Lambda}d^3 p_1 \int_{|{\bf p}_2|\le \Lambda}d^3 p_2  \nonumber\\
&& \quad \frac{E_\lambda({\bf p}_1) E_\lambda({\bf p}_2)+{\bf p}_1\cdot{\bf p}_2-m^2_\lambda}{E_\lambda({\bf p}_1) E_\lambda({\bf p}_2)} e^{i({\bf p}_1+{\bf p}_2)\cdot({\bf x}-{\bf y})}\,.
\label{a10}
\end{eqnarray}
In the last equality, we have used the relations
\begin{eqnarray}
\sum_{s} u_{\lambda,s,a}({\bf p})u^\dagger_{\lambda,s,a'}({\bf p})&=&\left(\frac{E_\lambda({\bf p}) + {\boldsymbol \alpha}\cdot {\bf p} + \beta m_\lambda}{2m_\lambda}\right)_{a a'}\,,\nonumber\\
\sum_{s} v_{\lambda,s,a}({\bf p})v^\dagger_{\lambda,s,a'}({\bf p})&=&\left(\frac{E_\lambda({\bf p}) + {\boldsymbol \alpha}\cdot {\bf p} - \beta m_\lambda}{2m_\lambda}\right)_{aa'}\,.
\label{a11}
\end{eqnarray}
So finally we obtain $\langle 0|\widehat{F}^2_d(B)|0\rangle=C_{I}+C_{II} = n_0(B)^2 + C_{II}$, so that $\left(\Delta_0 F_d(B)\right)^2 = \langle 0|\widehat{F}^2_d(B)|0\rangle - n_0(B)^2 = C_{II}$. 

We now evaluate this expression for a spherical region $B$, which we now denote by $B_s$, with radius $b$ and centered around ${\bf z}$. We are interested in radii $b$ which are of the order of the atomic distance $10^{-10}{\textrm{m}}=1{\textrm{\AA}}$ or larger.

In order to simplify the equations we will make the replacement
\begin{equation}
\int_{B_s} d^3x \to \sqrt{\frac{2}{9\pi}} \int d^3x e^{-\frac{|{\bf x}-{\bf z}|^2}{2 b^2}} \,,
\label{a12}
\end{equation}
where the factor $\sqrt{2/9\pi}$ is a normalization constant. In this way we get, after Gaussian integration over ${\bf x}$ and ${\bf y}$, and after change of integration variables $({\bf p}_1,{\bf p}_2) \to ({\bf p},{\bf q})$ where ${\bf p}_1={\bf p}$ and ${\bf p}_2= -{\bf p} - {\bf q}$, that{\footnote{The symbol `$\approx$' is used for `approximately equal as'. The symbol `$\sim$' will be used for `of the same order as'.}}
\begin{multline}
\left(\Delta_0 F_d(B_s)\right)^2 \approx  \frac{b^6}{36\pi^3} \sum_{\lambda}  \int_{|{\bf p}|\le \Lambda}d^3 p \int d^3 q e^{-b^2|{\bf q}|^2} \\
\times \frac{E_\lambda({\bf p}) E_\lambda({\bf p}+{\bf q})-{\bf p}\cdot({\bf p}+{\bf q})-m^2_\lambda}{E_\lambda({\bf p}) E_\lambda({\bf p}+{\bf q})}  \,.
\label{a12.1}
\end{multline}
We wrote an approximation sign instead of an equality sign because we dropped the integration bound $|{\bf q} + {\bf p}| \le \Lambda$ for ${\bf q}$. This can be done because of the following reason. The main contribution to the integral comes from momenta ${\bf q}$ for which $|{\bf q}|  \lesssim 1/b$, hence the bound $|{\bf q} + {\bf p}| \le \Lambda$ only plays a role when $|{\bf p}| \le \Lambda -1/b$. Since $ 1/b \ll\Lambda$ and since the integration domain of ${\bf p}$ is given by $|{\bf p}|\le \Lambda$, the removal of the bound $|{\bf q} + {\bf p}| \le \Lambda$ for ${\bf q}$ in the integral forms a good approximation. (Actually, more precisely, the right hand side forms of \eqref{a12.1} forms an upper bound for $\left(\Delta_0 F_d(B_s)\right)^2$, since the integrand is strictly positive, as can easily be checked.) We can further integrate this by passing to spherical coordinates, $({\bf p}) \to (p,\phi_p,\theta_p)$ and $({\bf q}) \to (q,\phi_q,\theta_q)$, where we choose $\theta_q$ to be the angle between ${\bf p}$ and ${\bf q}$. We obtain
\begin{eqnarray}
\left(\Delta_0 F_d(B_s)\right)^2 &\approx& \frac{2b^6}{9\pi^2} \sum_{\lambda} \int^{\Lambda}_0 dp \int^{+\infty}_{0} dq e^{-b^2q^2} \left( I_\lambda(p,q) + I_\lambda(p,-q) \right) \nonumber\\
&=& \frac{2b^6}{9\pi^2} \sum_{\lambda} \int^{\Lambda}_0 dp \int^{+\infty}_{-\infty} dq e^{-b^2q^2} I_\lambda(p,q)\,,
\label{a12.2}
\end{eqnarray}
where 
\begin{equation}
I_\lambda(p,q)=pq\left(pq - \frac{2}{3}E_\lambda(p+q) E_\lambda(p) + \frac{E_\lambda(p+q)}{3E_\lambda(p)} (q^2 - pq) \right)
\label{a14}
\end{equation}
and $E_\lambda(p)=\sqrt{p^2+m^2_\lambda}$. Note that $\left(\Delta_0 F_d(B_s)\right)^2$ is independent of the centre ${\bf z}$ of the spherical region $B_s$, which was to be expected since the vacuum $|0\ra$ is translation invariant.

We see that $\left(\Delta_0 F_d(B_s)\right)^2$ is the sum of contributions of the different types of particles, i.e.
\begin{equation}
\left(\Delta_0 F_d(B_s)\right)^2 = \sum_\lambda \left(\Delta_0 F_d(B_s)\right)^2_\lambda\,, 
\label{a14.1}
\end{equation}
where
\begin{equation}
 \left(\Delta_0 F_d(B_s)\right)^2_\lambda = \frac{2b^6}{9\pi^2}  \int^{\Lambda}_0 dp \int^{+\infty}_{-\infty} dq e^{-b^2q^2} I_\lambda(p,q)\,. 
\label{a14.2}
\end{equation}
In order to evaluate the integrals in $\left(\Delta_0 F_d(B_s)\right)^2_\lambda$, we consider the following separate cases:
\begin{itemize}
\item
{\bf Case 1:} $b \gg 1/m_\lambda$
\item
{\bf Case 2:} $b \leq 1/m_\lambda$
\item
{\bf Case 3:} Otherwise, i.e.\ $\left(b \geq 1/m_\lambda \right) \wedge \left(\lnot( b \gg 1/m_\lambda)\right)$
\end{itemize}
Remember that we are interested in radii $b$ which are of the order of the atomic distance $10^{-10}{\textrm{m}}=1{\textrm{\AA}}$ or larger. Therefore, the charged fermions, i.e.\ all types of fermions except neutrinos, always fall under Case 1. This is because the charged fermion with the largest Compton wavelength $1/m_\lambda$ is the electron, for which $1/m_\lambda \sim 10^{-12}{\textrm{m}}$. Cases 2 and 3 apply to neutrinos.

\subsection*{Case 1: $b \gg 1/m_\lambda$}
The main contribution to the integral comes from momenta $|q|$ for which $|q| \lesssim 1/b$. Since we also assume that the radius $b$ of the region $B_s$ is much bigger than the Compton wavelength, we have that the main contribution to the integral in $\left(\Delta_0 F_d(B_s)\right)^2_\lambda$ comes from momenta $|q|$ for which $|q| \ll m_\lambda$. By performing a Taylor expansion of $E_\lambda(p+q)$ around $q=0$, we find up to third order in $|q| / m_\lambda$ that  
\begin{equation}
I_\lambda(p,q) \approx -\frac{2}{3}E_\lambda(p)^2pq + \frac{1}{6E_\lambda(p)^4}(2p^4q^4 + 3p^2q^4m^2_\lambda)\,.
\label{a15}
\end{equation}
The first term will not contribute to the integral in $\left(\Delta_0 F_d(B_s)\right)^2_{\lambda,\Lambda}$ since it is odd in $q$. Doing the integration over the second term yields
\begin{equation}
\left(\Delta_0 F_d(B_s)\right)^2_\lambda \approx  \frac{1}{18\pi^{3/2}}  \Lambda b\,.
\label{a16}
\end{equation}
In this expression, we have dropped the terms of the order $bm_\lambda$ and lower since $m_\lambda \ll \Lambda$.

\subsection*{Case 2: $b \leq 1/m_\lambda$}
In order to evaluate $\left(\Delta_0 F_d(B_s)\right)^2_\lambda$ this time we write
\begin{equation}
 \left(\Delta_0 F_d(B_s)\right)^2_\lambda = \left(\Delta_0 F_d(B_s)\right)^2_{\lambda,\alpha} + \left(\Delta_0 F_d(B_s)\right)^2_{\lambda,\Lambda} \,, 
\label{a17}
\end{equation}
where
\begin{eqnarray}
\left(\Delta_0 F_d(B_s)\right)^2_{\lambda,\alpha} &=& \frac{2b^6}{9\pi^2} \int^{\alpha}_0 dp \int^{+\infty}_{-\infty} dq e^{-b^2q^2} I_\lambda(p,q)\,,\nonumber\\
\left(\Delta_0 F_d(B_s)\right)^2_{\lambda,\Lambda} &=& \frac{2b^6}{9\pi^2} \int^{\Lambda}_\alpha dp \int^{+\infty}_{-\infty} dq e^{-b^2q^2} I_\lambda(p,q)\,.
\label{a18}
\end{eqnarray}
We take $\alpha$ such that $\Lambda \gg \alpha \gg 1/b$. Together with $b \le 1/m_\lambda$ we have that $\alpha \gg 1/b \ge m_\lambda $.

Let us first evaluate $\left(\Delta_0 F_d(B_s)\right)^2_{\lambda,\Lambda}$. Since the main contribution to the integrals comes from momenta $|q|$ for which $|q| \lesssim  1/b$, the particular choice for $\alpha$ implies that the main contribution to $\left(\Delta_0 F_d(B_s)\right)^2_{\lambda,\Lambda}$ comes from momenta $p$ for which $p \gg m_\lambda$ and $|p + q| \gg m_\lambda$. For $p \gg m_\lambda$, $|p  + q| \gg m_\lambda$ and $(p,q) \in [0,\alpha] \times [-1/b,1/b]$, we have that
\begin{equation}
I_\lambda(p,q) \approx \frac{1}{3} q^4 - \frac{2}{3}p^3q \,.
\label{a19}
\end{equation}
The second term will not contribute to the integral in $\left(\Delta_0 F_d(B_s)\right)^2_{\lambda,\Lambda}$ since it is odd in $q$. Doing the integration over the first term yields
\begin{equation}
\left(\Delta_0 F_d(B_s)\right)^2_{\lambda,\Lambda}   \approx  \frac{1}{18\pi^{3/2}}  \Lambda b  \,.
\label{a20}
\end{equation}
This is the same expression as in ({\ref{a16}}).

We now show that $\left(\Delta_0 F_d(B_s)\right)^2_{\lambda,\alpha} \ll \left(\Delta_0 F_d(B_s)\right)^2_{\lambda,\Lambda}$. In order to do so, first note that for $(p,q) \in [0,\alpha] \times [-1/b,1/b]$ we have that 
\begin{eqnarray}
\left| I_\lambda(p,q) \right| &\le& p^2 q^2 + \frac{2}{3}p|q|  E_\lambda(p+q) E_\lambda(p) + \frac{E_\lambda(p+q)}{3E_\lambda(p)} \left( |q|^3p + |q|^2p^2  \right) \nonumber\\
&\le& p^2 q^2 + \frac{2}{3}p|q|  E_\lambda(p+q) E_\lambda(p) + \frac{1}{3} E_\lambda(p+q) \left( |q|^3 + |q|^2p  \right) \nonumber\\
 &\le& \frac{\alpha^2}{b^2} + \frac{2}{3} \frac{\alpha}{b}  \sqrt{\left(\alpha + \frac{1}{b} \right)^2 +m^2_\lambda } \sqrt{\alpha^2 + m^2_\lambda }  \nonumber\\
&& + \frac{1}{3} \sqrt{\left(\alpha + \frac{1}{b} \right)^2 +m^2_\lambda }\left( \frac{1}{b^3} + \frac{\alpha}{b^2} \right) \nonumber\\
 &\approx&  \frac{2}{3} \frac{\alpha^3}{b} \,,
\end{eqnarray}
where in the last equation we have used the fact that $\alpha \gg m_\lambda$ and $\alpha \gg 1/b$. As a result we have
\begin{equation}
\left(\Delta_0 F_d(B_s)\right)^2_{\lambda,\alpha} \lesssim  \frac{2}{9\pi^{3/2}} \alpha^4b^4    \,.
\label{a20.1}
\end{equation}
Now we check that $\left(\Delta_0 F_d(B_s)\right)^2_{\lambda,\alpha}$ is negligible compared to $\left(\Delta_0 F_d(B_s)\right)^2_{\lambda,\Lambda}$. To this end assume $\alpha \sim 10^3/b$ ($\alpha$ was chosen such that $\Lambda \gg \alpha \gg 1/b \ge m_\lambda$). We then have that $\left(\Delta_0 F_d(B_s)\right)^2_{\lambda,\alpha} \lesssim 2 \cdot 10^{12} / 9\pi^{3/2} \sim 10^{10}$. If we assume the cut-off to be of the order of the Planck mass, i.e.\ $\Lambda \sim 10^{35}/{\textrm{m}}$, then $\left(\Delta_0 F_d(B_s)\right)^2_{\lambda,\alpha} $ is clearly much smaller than $\left(\Delta_0 F_d(B_s)\right)^2_{\lambda,\Lambda} \approx  b \Lambda/18\pi^{3/2} \sim 10^{-2} \Lambda b \sim 10^{33} b/{\textrm{m}}$ for $b \ge 10^{-10} {\textrm{m}} =1{\textrm{\AA}} $. Hence we obtain
\begin{equation}
\left(\Delta_0 F_d(B_s)\right)^2_\lambda  \approx \left(\Delta_0 F_d(B_s)\right)^2_{\lambda,\Lambda}  \approx  \frac{1}{18\pi^{3/2}}  \Lambda b \,.
\label{a21}
\end{equation}

\subsection*{Case 3: Otherwise, i.e.\ $\left(b \geq 1/m_\lambda \right) \wedge \left(\lnot( b \gg 1/m_\lambda)\right)$}
This case corresponds to $b$ larger than the Compton wavelength of the fermion but not too much larger either. Therefore we take $b = 10^{\beta}/m_\lambda$, where we take $\beta \in [0,3]$.

In this case, we use the same technique as in the previous case. We split up the integral like in equation (\ref{a17}). We choose $\Lambda \gg \alpha \gg m_\lambda$. Together with the conditions on $b$, we have then that $\alpha \gg m_\lambda \ge 1/b \ge 10^{-3}m_\lambda$. Just as in the previous case, it follows that the main contribution to the integral in $\left(\Delta_0 F_d(B_s)\right)^2_{\lambda,\Lambda}$ comes from momenta $p$ for which $p \gg m_\lambda$ and $|p + q| \gg m_\lambda$. Therefore we have the same expression for $\left(\Delta_0 F_d(B_s)\right)^2_{\lambda,\Lambda}$ and the same bound for $\left(\Delta_0 F_d(B_s)\right)^2_{\lambda,\alpha}$. 

We also have that $\left(\Delta_0 F_d(B_s)\right)^2_\lambda$ is dominated by $\left(\Delta_0 F_d(B_s)\right)^2_{\lambda,\Lambda}$. To see this, consider the ratio $\varepsilon = \left(\Delta_0 F_d(B_s)\right)^2_{\lambda,\alpha} / \left(\Delta_0 F_d(B_s)\right)^2_{\lambda,\Lambda} \approx 4 \alpha^4b^3 / \Lambda$. If $\varepsilon \ll 1$ then $\left(\Delta_0 F_d(B_s)\right)^2_\lambda$ will be dominated by $\left(\Delta_0 F_d(B_s)\right)^2_{\lambda,\Lambda}$. Now, since $\Lambda \gg \alpha \gg m_\lambda \ge 1/b$, we can take $\alpha \sim 10^3 m_\lambda$. In this way, we get $\varepsilon \lesssim 4 \cdot 10^{-23 +3 \beta} m_\lambda {\textrm{m}}$. Since the Case 3 only applies to neutrinos, we can find an upper bound for $\varepsilon$ by substituting $m_\lambda$ by the upper bound for the neutrino mass. Cosmological considerations imply that the sum of the masses of the different neutrino flavours is smaller than $2{\textrm{eV}} \simeq 4\cdot 10^{6}/{\textrm{m}}$ \cite[p.\ 474]{yao06}. Therefore an upper bound for $\varepsilon$ is given by $1.6 \cdot 10^{-16 +3 \beta}$. Clearly this upper bound is still much smaller than 1 for $\beta \in [0,3]$. Therefore we have, just as in the previous cases, that 
\begin{equation}
\left(\Delta_0 F_d(B_s)\right)^2_{\lambda,\Lambda} \approx  \frac{1}{18\pi^{3/2}}  \Lambda b   \,.
\label{a21.1}
\end{equation}

\subsection*{Summary}
Combining the above results we find that 
\begin{equation}
\left(\Delta_0 F_d(B_s)\right)^2 = \sum_\lambda \left(\Delta_0 F_d(B_s)\right)^2_\lambda \approx  \sum_\lambda \frac{1}{18\pi^{3/2}}  \Lambda b  = \frac{4 }{3\pi^{3/2}} \Lambda b=  \left(\frac{4}{3} \right)^{2/3} \frac{1}{\pi^{11/6}} \Lambda V(B_s)^{1/3}\,, 
\label{a22}
\end{equation}
where $V(B_s)=4\pi b^3 /3$ is the volume of the spherical region $B_s$. The standard deviation now reads
\begin{equation}
\Delta_0 F_d(B_s)  \approx \left(\frac{4}{3} \right)^{1/3} \frac{1}{\pi^{11/12}} \Lambda^{1/2} V(B_s)^{1/6}  \approx 0.39 \Lambda^{1/2} V(B_s)^{1/6} \,.
\label{a23}
\end{equation}

\section{Expectation value and standard deviation of the fermion number for a macroscopic state $|\varphi \ra$}\label{expectationvaluemacroscopic}
In this appendix, we calculate the expectation value and the standard deviation for the fermion number in a region $B$ which is of the order of the atomic distance $10^{-10}{\textrm{m}}=1{\textrm{\AA}}$ or larger, for the state $| \vp \rangle$ given in \eqref{30.0001}, which represents $m$ particles (and no anti-particles) on top of the vacuum $|0\ra$ that are approximately localized within $B$. Before we calculate the expectation value and the standard deviation, we first introduce some notation and useful identities, as well as some approximations. 

\subsection{Some notation and useful identities}
We first introduce the operator
\begin{eqnarray}
{\widehat C}^{\dagger}_\varphi &=& \frac{1}{\sqrt{m!}} \sum_{\lambda_1,a_1,\dots,\lambda_m,a_m} \int d^3 x_1 \dots d^3 x_m \varphi_{\lambda_1,a_1,\dots,\lambda_m,a_m}({\bf x}_1,\dots,{\bf x}_m) \nonumber\\
&& \qquad \qquad \qquad   \qquad \qquad \qquad  \qquad \qquad  \qquad  \times {\widehat \psi}^{\dagger}_{\lambda_1,a_1}({\bf x}_1) \dots  {\widehat \psi}^{\dagger}_{\lambda_m,a_m}({\bf x}_m) \nonumber\\
&=& \frac{1}{\sqrt{m!}} \sum_{\lambda_1,s_1,\dots,\lambda_m,s_m} \int_{|{\bf p}_1| \le \Lambda} d^3 p_1 \dots \int_{|{\bf p}_m| \le \Lambda} d^3 p_m {\widetilde \varphi}_{\lambda_1,s_1,\dots,\lambda_m,s_m}({\bf p}_1,\dots,{\bf p}_m)  \nonumber\\
&& \qquad \qquad \qquad   \qquad \qquad \qquad  \qquad \qquad   \qquad \times {\widehat c}^{\dagger}_{\lambda_1,s_1}({\bf p}_1) \dots  {\widehat c}^{\dagger}_{\lambda_m,s_m}({\bf p}_m)\,,
\label{a57}
\end{eqnarray}
for which ${\widehat C}^{\dagger}_\varphi \left| 0 \right\ra  = \left| \varphi \right\ra$. 

We have that 
\begin{equation}
\left[ \widehat{F}_d(B), {\widehat C}^{\dagger}_\varphi \right]  = m {\widehat C}^{\dagger}_{\varphi}(B)  \,,
\label{a58}
\end{equation}
where
\begin{multline}
{\widehat C}^{\dagger}_{\varphi}(B) = \frac{1}{\sqrt{m!}} \sum_{\lambda_1,a_1,\dots,\lambda_m,a_m} \int_B d^3 x_1 \int d^3 x_2 \dots d^3 x_m \varphi_{\lambda_1,a_1,\dots,\lambda_m,a_m}({\bf x}_1,\dots,{\bf x}_m) \\
\times {\widehat \psi}^{\dagger}_{\lambda_1,a_1}({\bf x}_1) \dots  {\widehat \psi}^{\dagger}_{\lambda_m,a_m}({\bf x}_m) \,.
\label{a59}
\end{multline}
This follows from
\begin{multline}
\left[ \widehat{F}_d(B), {\widehat \psi}^{\dagger}_{\lambda_1,a_1}({\bf x}_1) \dots  {\widehat \psi}^{\dagger}_{\lambda_m,a_m}({\bf x}_m) \right]  \\
=\sum^m_{i=1} \int_B d^3 x \delta^{(\Lambda)}({\bf x}-{\bf x}_i){\widehat \psi}^{\dagger}_{\lambda_1,a_1}({\bf x}_1) \dots \left[ {\widehat \psi}^{\dagger}_{\lambda_i,a_i}({\bf x}_i)  \to  {\widehat \psi}^{\dagger}_{\lambda_i,a_i}({\bf x})  \right] \dots  {\widehat \psi}^{\dagger}_{\lambda_m,a_m}({\bf x}_m) \,,
\label{a60}
\end{multline}
where $\left[ {\widehat \psi}^{\dagger}_{\lambda_i,a_i}({\bf x}_i)  \to  {\widehat \psi}^{\dagger}_{\lambda_i,a_i}({\bf x})  \right] $ means that $ {\widehat \psi}^{\dagger}_{\lambda_i,a_i}({\bf x}_i)$ is replaced by ${\widehat \psi}^{\dagger}_{\lambda_i,a_i}({\bf x})$. 

We further have that
\begin{equation}
\left[ \widehat{F}_d(B), {\widehat C}^{\dagger}_\varphi(B) \right]  = m {\widehat {\widetilde C}}^{\dagger}_{\varphi}(B)  \,,
\label{a61}
\end{equation}
where
\begin{multline}
{\widehat {\widetilde C}}^{\dagger}_{\varphi}(B) =  \frac{1}{\sqrt{m!}} \sum_{\lambda_1,a_1,\dots,\lambda_m,a_m} \int_B d^3 x_1 \int d^3 x_2 \dots d^3 x_m \varphi_{\lambda_1,a_1,\dots,\lambda_m,a_m}({\bf x}_1,\dots,{\bf x}_m) \\
\times  \left( \frac{(m-1)}{m} {\widehat \psi}^{\dagger}_{\lambda_1,a_1}({\bf x}_1) +  \frac{1}{m} \int_B d^3 x {\widehat \psi}^{\dagger}_{\lambda_1,a_1}({\bf x}) \delta^{(\Lambda)}({\bf x}-{\bf x}_1)\right) {\widehat \psi}^{\dagger}_{\lambda_2,a_2}({\bf x}_2) \dots  {\widehat \psi}^{\dagger}_{\lambda_m,a_m}({\bf x}_m) \,.
\label{a62}
\end{multline}

Other useful relations are
\begin{eqnarray}
&{\widehat C}_\varphi \left| \varphi \right\ra =  {\widehat C}_\varphi {\widehat C}^{\dagger}_\varphi \left| 0 \right\ra  =  \left| 0 \right\ra \,, & \nonumber\\
&{\widehat C}_{\varphi} {\widehat C}^{\dagger}_{\varphi}(B) | 0 \ra =  \la 0 | {\widehat C}_{\varphi} {\widehat C}^{\dagger}_{\varphi}(B) | 0 \ra  | 0 \ra=  \la \varphi | 
{\widehat C}^{\dagger}_{\varphi}(B) | 0 \ra  | 0 \ra \,. &
\label{a63}
\end{eqnarray}
The first relation can be derived as follows. The state ${\widehat C}_\varphi \left| \varphi \right\ra = {\widehat C}_{\varphi} {\widehat C}^{\dagger}_\varphi  | 0 \ra$ contains the product ${\widehat c} \dots {\widehat c} {\widehat c}^\dagger \dots {\widehat c}^\dagger$ of $m$ annihilation operators and $m$ creation operators acting on the state $| 0 \ra$. Moving the annihilation operators to the right, we find that ${\widehat C}_{\varphi} \left| \varphi \right\ra \sim | 0 \ra$ and hence, by using the resolution of the identity, we have $ {\widehat C}_{\varphi} \left| \varphi \right\ra=  | 0 \ra \la 0 |{\widehat C}_{\varphi} | \varphi \ra  = | 0 \ra \la  \varphi | \varphi \ra =  | 0 \ra$. The second relation can be derived completely analogously.

\subsection{Some approximations}
We now present some approximations which we will use in the calculation of the expectation value and the standard deviation. The approximations have to do with the fact that the wavefunction $\varphi_{\lambda_1,a_1,\dots,\lambda_m,a_m}({\bf x}_1,\dots,{\bf x}_m)$ only has most, and not all, of its support in the region $B$ (as explained in Section \ref{typicalbeableconfigurations}). 

First, we have
\begin{eqnarray}
\la \varphi | {\widehat C}^\dagger_{\varphi}(B) | 0 \ra &=& \frac{1}{\sqrt{m!}} \sum_{\lambda_1,a_1,\dots,\lambda_m,a_m} \int_B d^3 x_1 \int d^3 x_2 \dots d^3 x_m \varphi_{\lambda_1,a_1,\dots,\lambda_m,a_m}({\bf x}_1,\dots,{\bf x}_m) \nonumber\\
&& \qquad \qquad \qquad \qquad  \qquad \qquad \qquad \times \la \varphi |  {\widehat \psi}^{\dagger}_{\lambda_1,a_1}({\bf x}_1) \dots  {\widehat \psi}^{\dagger}_{\lambda_m,a_m}({\bf x}_m)  \left| 0 \right\ra \nonumber\\
&=& \sum_{\lambda_1,a_1,\dots,\lambda_m,a_m}  \int_B d^3 x_1 \int d^3 x_2 \dots d^3 x_m |\varphi_{\lambda_1,a_1,\dots,\lambda_m,a_m}({\bf x}_1,\dots,{\bf x}_m)|^2 \nonumber\\
&\approx& 1\,.
\label{a64}
\end{eqnarray}
The approximation in the last line stems from the fact that most of the support of the wavefunction $\varphi_{\lambda_1,a_1,\dots,\lambda_m,a_m}({\bf x}_1,\dots,{\bf x}_m)$ is within the region $B^m \subset {\mathbb R}^{3m}$. For similar reasons we have
\be
\la 0 | {\widehat C}_{\varphi}(B) {\widehat C}^\dagger_{\varphi}(B) | 0 \ra \approx 1\,.
\label{a65}
\ee

We also have
\begin{align}
\la \varphi | {\widehat {\widetilde C}}^\dagger_{\varphi}(B) | 0 \ra =& \sum_{\lambda_1,a_1,\dots,\lambda_m,a_m} \int_B d^3 x_1 \int d^3 x_2 \dots d^3 x_m \varphi^*_{\lambda_1,a_1,\dots,\lambda_m,a_m}({\bf x}_1,\dots,{\bf x}_m) \nonumber\\
& \qquad \times \bigg( \frac{(m-1)}{m} \varphi_{\lambda_1,a_1,\dots,\lambda_m,a_m}({\bf x}_1,\dots,{\bf x}_m) \nonumber\\
& \qquad \qquad  +  \frac{1}{m} \int_B d^3 x  \delta^{(\Lambda)}({\bf x}-{\bf x}_1)  \varphi_{\lambda_1,a_1,\dots,\lambda_m,a_m}({\bf x},{\bf x}_2 \dots,{\bf x}_m) \bigg) \nonumber\\
\approx&  \frac{m-1}{m} + \frac{1}{m} =  1 \,.
\label{a66}
\end{align} 
Here the approximation stems not only from the fact that most of the support of the wavefunction is within $B^m$, but also from the fact that most of the support of the function $\delta^{(\Lambda)}$ is located within a region of the order of the Planck length $1/\Lambda$, so that we can approximate the expression by replacing it with the ordinary $\delta$-function.

We further have that
\begin{align}
\la \varphi | \widehat{F}_d(B)  {\widehat C}^\dagger_{\varphi}(B) | 0 \ra &= \la 0 | {\widehat C}_{\varphi} \widehat{F}_d(B)  {\widehat C}^\dagger_{\varphi}(B) | 0 \ra \nonumber\\
&= \la 0 | \left[ {\widehat C}_{\varphi}, \widehat{F}_d(B) \right]  {\widehat C}^\dagger_{\varphi}(B) | 0 \ra + \la 0 | \widehat{F}_d(B)  {\widehat C}_{\varphi}  {\widehat C}^\dagger_{\varphi}(B) | 0 \ra  \nonumber\\
&= m \la 0 |{\widehat C}_{\varphi}(B)  {\widehat C}^\dagger_{\varphi}(B) | 0 \ra +  \la \varphi| {\widehat C}^\dagger_{\varphi}(B) | 0 \ra \la 0 | \widehat{F}_d(B) | 0 \ra  \nonumber\\
&\approx m + n_0(B)\,,
\label{a67}
\end{align} 
where in the last line we respectively used \eqref{a65} and \eqref{a64}. With the help of this relation and \eqref{a66} we have that
\begin{align}
\la \varphi |  {\widehat C}^\dagger_{\varphi}(B) \widehat{F}_d(B)  | 0 \ra &= \la \varphi | \widehat{F}_d(B)  {\widehat C}^\dagger_{\varphi}(B)  | 0 \ra - \la \varphi | \left[\widehat{F}_d(B) ,{\widehat C}^\dagger_{\varphi}(B)\right]  | 0 \ra \nonumber\\
&= \la \varphi | \widehat{F}_d(B)  {\widehat C}^\dagger_{\varphi}(B)  | 0 \ra -m \la \varphi |{\widehat {\widetilde C}}^\dagger_{\varphi}(B) | 0 \ra \nonumber\\
&\approx m+n_0(B) - m = n_0(B)\,.
\label{a68}
\end{align}

\subsection{Expectation value and standard deviation}
Using the approximations, we obtain the expectation value
\begin{eqnarray}
\left\la \varphi \right| \widehat{F}_d(B) \left| \varphi  \right\ra &=& \left\la \varphi  \right| \widehat{F}_d(B) {\widehat C}^\dagger_\varphi \left| 0 \right\ra \nonumber\\
&=& \left\la \varphi  \right| \left[ \widehat{F}_d(B), {\widehat C}^\dagger_\varphi \right] \left| 0 \right\ra + \left\la \varphi  \right|  {\widehat C}^\dagger_\varphi \widehat{F}_d(B) \left| 0 \right\ra \nonumber\\
&=& m \la \varphi | {\widehat C}^\dagger_{\varphi}(B) | 0 \ra  +n_0(B)\nonumber\\
&\approx& m +n_0(B)
\label{a69}
\end{eqnarray}
and with
\begin{eqnarray}
\left\la \varphi \right| \widehat{F}_d(B)^2 \left| \varphi  \right\ra &=& \left\la \varphi  \right| \widehat{F}_d(B)^2 {\widehat C}^\dagger_\varphi \left| 0 \right\ra \nonumber\\
&=& \left\la \varphi  \right|\widehat{F}_d(B) \left[ \widehat{F}_d(B), {\widehat C}^\dagger_\varphi \right] \left| 0 \right\ra + \left\la \varphi  \right| \widehat{F}_d(B) {\widehat C}^\dagger_\varphi \widehat{F}_d(B) \left| 0 \right\ra \nonumber\\
&=& m \la \varphi | \widehat{F}_d(B)  {\widehat C}^\dagger_{\varphi}(B) | 0 \ra  +  \left\la \varphi  \right| \left[ \widehat{F}_d(B), {\widehat C}^\dagger_\varphi \right] \widehat{F}_d(B) \left| 0 \right\ra  + \left\la \varphi  \right| {\widehat C}^\dagger_\varphi \widehat{F}_d(B)^2  \left| 0 \right\ra \nonumber\\
&=& m \la \varphi | \widehat{F}_d(B)  {\widehat C}^\dagger_{\varphi}(B) | 0 \ra + m \left\la \varphi  \right| {\widehat C}^\dagger_{\varphi}(B)\widehat{F}_d(B) | 0 \ra +  \left\la 0 \right|\widehat{F}_d(B)^2 \left| 0 \right\ra \nonumber\\
&\approx& m^2 +2mn_0(B) + \left\la 0 \right|\widehat{F}_d(B)^2 \left| 0 \right\ra \,,
\label{a70}
\end{eqnarray}
we find the standard deviation
\begin{equation}
\Delta_\varphi F_d(B) =  \sqrt{\left\la \varphi \right| \widehat{F}_d(B)^2 \left| \varphi \right\ra - \left\la \varphi \right| \widehat{F}_d(B) \left| \varphi \right\ra^2 } \approx \Delta_0 F_d(B) \,.
\label{a71}
\end{equation}

Completely analogously one can calculate the expectation value $\left\la \varphi \right| \widehat{F}_d({\tilde B}) \left| \varphi  \right\ra$ and standard deviation $\Delta_\varphi F_d({\tilde B})$ for a region ${\tilde B} \subset {\mathbb R}^{3} \setminus B$. The result is:
\begin{equation}
\left\la \varphi \right| \widehat{F}_d({\tilde B}) \left| \varphi  \right\ra \approx n_0({\tilde B})  \,,\qquad \Delta_\varphi F_d({\tilde B} )\approx \Delta_0 F_d({\tilde B} ) \,.
\label{a72}
\end{equation}

\end{document}